\documentclass[12pt,twoside]{article}

\newif\ifTIKZinline
\TIKZinlinetrue
\TIKZinlinefalse

\usepackage{amssymb}
\usepackage{amsmath}
\usepackage{tikz}
\usetikzlibrary{arrows.meta,spy,backgrounds} %

\usepackage{pgfplots}
\usepackage{pgfplotstable}
\usepgfplotslibrary{units}
\pgfplotsset{compat=newest} %

\usepackage{diagbox} %
\usepackage{makecell} %
\usepackage{enumitem} %

\usepackage[utf8]{inputenc}

\usepackage[hypertex]{hyperref} %
\usepackage[british]{babel} %

\usepackage{float} %

\newfloat{algorithm}{tbp}{loa} %
\floatname{algorithm}{Algorithm}
\floatstyle{ruled} %
\restylefloat{algorithm}

\usepackage{caption}%
\DeclareCaptionStyle{mystyle}%
[labelfont=bf,justification=centering]%
{labelfont=bf,justification=centerlast}%

\usepackage{my}
\usepackage[commentColor=commentGreen,noEnd=false]{algpseudocodex} %

\algrenewcommand\alglinenumber[1]{\scriptsize #1:} %
\algnewcommand{\IIf}[1]{\State\algorithmicif\ #1\ \algorithmicthen} %
\algnewcommand{\ElseIIf}[1]{\\ \algorithmicelse\ \algorithmicif\ #1} %
\algnewcommand{\ElseI}[1]{\State\algorithmicelse\ \ \ \ #1} %
\algnewcommand{\EndIIf}{\unskip\ }
\algnewcommand{\IFor}[1]{\State\algorithmicfor\ #1} %
\algnewcommand{\NoEndIf}{\unskip\algorithmicend}

\algnewcommand{\WWhile}[1]{\State\algorithmicwhile\ #1\ \algorithmicdo} %
\algnewcommand{\EndWWhile}{\unskip\ }

\algnewcommand{\IfThen}[3]{%
  \State \algorithmicif\ #1\ \algorithmicthen\ #2}
\algnewcommand{\IfThenElse}[3]{%
  \State \algorithmicif\ #1\ \algorithmicthen\ #2\ \algorithmicelse\ #3}
\def\monthword{\ifcase\month \or
  January\or February\or March\or April\or May\or June\or
  July\or August\or September\or October\or November\or December\fi}

\newcommand{\TikZ}{Ti\emph{k}Z}
\newcommand{\BibTeX}{B\textsc{ib}$\!$\TeX}
\begin{document}
\pagestyle{myheadings}
\markboth{T.Strutz: Initial Parameter Estimation, Trigonometric Function}{TECHP/2026/01}
\title{\vspace{-3em}\bf Initial Parameter Estimation for Non-Linear Optimization\\ -- Trigonometric Function -- }
\author{\copyright Tilo Strutz\\%
 Technical paper, 2026, \monthword, TECHP/2026/01\\%
version: \today \\%
 Coburg University\\%
Faculty of Electrical Engineering 
               and Computer Science
}
\date{\vspace{-5ex}} %
\maketitle
\begin{abstract}
Nonlinear optimisation techniques are commonly employed to minimise complex cost functions, with their effectiveness determined largely by the structure of the underlying error landscape. These methods require initial parameter values, and in the presence of multiple local minima, they are prone to becoming trapped in suboptimal regions. The likelihood of locating the global minimum increases substantially when the initialisation lies within its corresponding basin of attraction. Consequently, high-quality initial parameters are critical for successful optimisation.
This technical report outlines a new strategy for selecting suitable initial parameters for a trigonometric model and unevenly sampled data, ensuring that the optimisation procedure starts sufficiently close to the global minimum. 
The proposed parameter estimation approach is strictly NI-based, interpretable, and explainable. It targets at complicated cases which include: samples with strong random noise, samples with only few covered periods, and samples which cover only a fraction of one period.
Special attention is put on the frequency estimation.
It can be shown that an estimation of initial parameters with sufficient accuracy is possible down to a signal-noise-ratio of 1.4 dB at much lower computational costs than the Lomb-Scargle-periodogram method requires.
\end{abstract}

\section{Introduction}
\label{sec_intro}
Nonlinear optimization plays a central role in many scientific and engineering applications, where the goal is to minimize a cost function that measures the discrepancy between a model and observed data. Unlike linear problems, nonlinear optimization involves navigating an error landscape that often is highly complex and may contain numerous local minima as well as flat regions with very low gradients. This complexity introduces a significant challenge: standard optimization algorithms, such as gradient-based methods, can easily become trapped in a local minimum, failing to reach the global optimum.
The structure of the error landscape strongly influences the success of the optimization process. In particular, the probability of finding the global minimum is much higher when the algorithm starts in the valley that contains the global solution. If the initial parameters are chosen poorly, the optimization may begin in a region far from the desired basin of attraction, leading to slow convergence or convergence to an incorrect solution. Consequently, selecting good initial start parameters $\mathbf{a}=(a_1 \; \dots\; a_j \; \dots\; a_M)$ is not merely a convenience -- it is essential for reliable and efficient optimization.

The general advise for parameter estimation is to utilise the physical meaning of the model function. Based on the observed values $y_i$, typically some assumption can be made leading to a rough guess of $\mathbf{a}$.

This technical report focuses on strategies for determining suitable initial parameters for a trigonometric model function of the form:
\begin{align}\label{eq_cosFunc}
	f(x|\mathbf{a}) = y = a_1 + a_2\cdot \cos(a_3\cdot x + a_4) \qquad\mbox{with}\quad a_1,a_2,a_3,a_4\in \ReelSet
	\;,
\end{align}
as it is widely used in fields such as physics and engineering.
The parameters $a_j$ are unknown and have to be estimated from data. 

\begin{figure}
\hfil
	\includegraphics[scale=0.8]{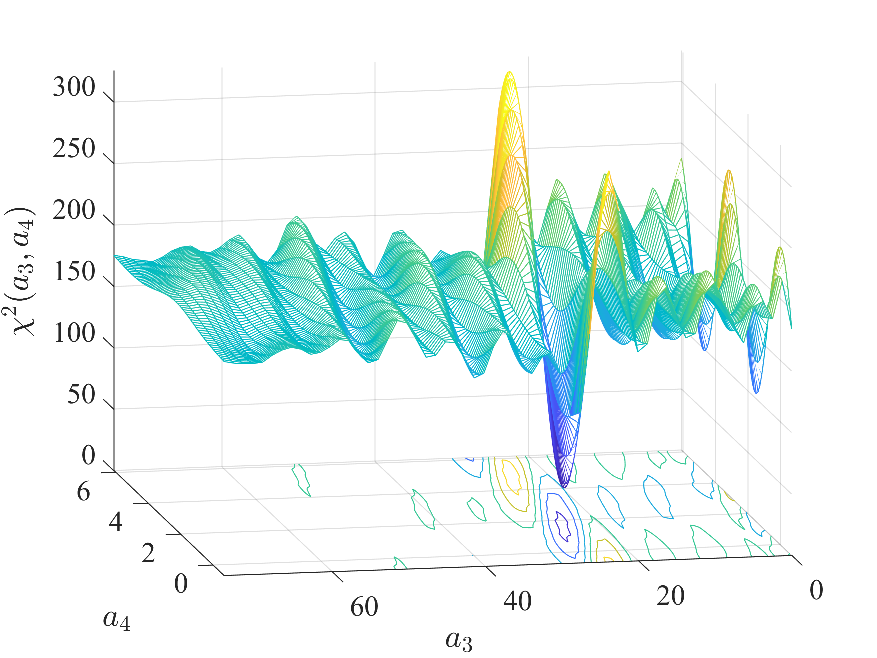}
	\caption{\label{fig_errorLandscape}Error landscape $\chi^2(\mathbf{a})$ of a trigonometric function as a function of frequency and phase (offset $a_1$ and amplitude $a_2$ are kept constant)}
\end{figure}

While estimating the offset $a_1$ and the amplitude $a_2$ is rather straightforward, finding a suitable start value for the frequency $f=a_3/(2\pi)$ is important because the periodicity of trigonometric functions causes many ambiguities. \Figu{fig_errorLandscape} shows the sum of squared differences between the observations and the predictions of a model function example dependent on the chosen parameters $a_3$ (frequency) and $a_4$ (phase): 
\begin{align}\label{eq_costFunction}
	\chi^2(\mathbf{a}) = \sum_{i=1}^N\left[y_i - f(x|\mathbf{a})\right]^2
	\;.
\end{align}
The global minimum can distinctly be seen at about $a_3=26, a_4 =1.5$.
However, at higher and lower values of $a_3$, there are several region separated by barriers from the valley containing the global minimum. 
If the optimization procedure would start behind such barrier, it becomes impossible to reach the wanted target.

The data to be fitted consist of pairs $(x_i;y_i)$ with  $i=1,2,\dots , N$, where $y_i$ is the observation measured in an $i$th experiment while using the condition $x_i$. Typically, the observations are subject to random errors, while the conditions $x_i$ are assumed to be error-free or, at least, the error is negligible compared to the errors in $y_i$.

As long as the conditions $x_i$ are equidistantly sampled, the frequency could be estimated using the discrete Fourier transform or the autocorrelation function. 
For unevenly sampled data, the Lomb-Scargle periodogram is the method of choice of detecting the dominant frequency, see \cite{Lomb76,VdP18} and cited papers within. It computes the spectral power for a plenty of possible frequency candidates and selects the frequency with the maximum response:
\begin{align}\label{eq_LombScargleP}
	p(\omega)&={\D\frac {1}{2}
\left[\frac {\left[ \sum\limits_{i}y_{i}\cos ( \omega x_{i}-\varphi )\right]^{2}}
                   {\sum\limits_{i}\cos ^{2} ( \omega x_{i}-\varphi )}+
			\frac {\left[ \sum\limits_{i}y_{i}\sin ( \omega x_{i}-\varphi )\right]^{2}}
			             {\sum\limits_{i}\sin ^{2} ( \omega x_{i}-\varphi )}\right]} \\
									\label{eq_LombScarglePhi}
	 \varphi &=\frac{1}{2}\arctan\left(\frac{\sum\limits_{i}\sin 2\omega x_{i}}
																					{\sum\limits_{i}\cos 2\omega x_{i}}\right)
	\qquad \mbox{with~} \omega = 2\pi\cdot f
	\;.
\end{align}
If nothing is known about the actual frequency, many different values from a dense grid must be tried out. This makes calculating the Lomb-Scargle periodogram quite time-consuming.

For determining the parameter vector $\hat{\mathbf{a}}$ of (\ref{eq_cosFunc}), a generally applicable and low-complexity approach is proposed which does not make any restriction with respect to the domain of definition of neither the parameters $a_j$ nor the observations $y_i$ or conditions $x_i$. The idea is to derive the parameter domains directly from the observations. The approach makes robust initial guesses and is functional for a broad range of measurement and sampling conditions. 
The proposed method is called `Fast Initial Parameter Estimation For Trigonometric functions' (FIPEFT).

Focus is put on the frequency estimation which can be achieved, for example, by examining the zero-crossings of the trigonometric function \cite{Dec85,Ked86,Fri94,Sad96,Gri12,Ang14,Men14,Bus19,Yur24}.
In contrast to earlier publications, the proposed method does not attempt to determine the exact frequency. 
Instead it is aiming at a fast procedure with sufficient accuracy allowing the final estimation by non-linear optimization techniques such as the Levenberg-Marquardt method \cite{Stru16}.  This includes signals which contain only a few or even fractions of periods.
Detailed algorithms are given enabling the reproducible research.

The effectiveness of the proposed approach is validated for a variety of test data in terms of robustness against measurement noise, sampling frequency, and the number of covered periods of the function. The frequency determination is compared to the results when using the Lomb-Scargle periodogram.

For better understanding, \Table{tab_notation} lists the symbols used throughout the text.
\begin{table}
	\caption{\label{tab_notation}Notation}
	\hfil
	\begin{tabular}{|l|l|}
		\hline
		variable			& meaning		\\
		\hline
		$a_j$					& parameter of model function		\\
		$\mathbf{a}$	& vector of parameters	\\
		$\hat{a}_j$		& estimated parameter 		\\
		$d_k$					& any distance			\\
		$d_{0,i}$			& good distance			\\
		$d_{{\rm s},j}$			& spurious distance			\\
		$d_{\rm ref}$	& reference distance (intermediate)			\\
		$d_{\rm typ}$	& typical distance (intermediate)			\\
		$d^*$					& best distance estimate (final)			\\
		$f$						& frequency value			\\
		$f_{\rm s}$					& sampling frequency			\\
		$N$						& number of data points $(x_i;y_i)$			\\
		$P$						& number of periods covered in data		\\
		$\mathbf{p}$	& vector of power values			\\
		$\sigma$			& standard deviation (of noise)			\\
		$T$						& $=1/f$ duration of one period			\\
		$x_i$					& independent variable of data to be processed			\\
		$x_k$					& $x$-value at which 	$y_i$ is maximal		\\
		$x_l$					& $x$-value at which 	$y_i$ is minimal		\\
		$y_i$					& observation, dependent variable of data			\\
		$yMax$				& $\max\limits_i(y_i)$		\\
		$yMin$				& $\min\limits_i(y_i)$		\\
		$yMax2$				& $\max\limits_i(y_i)$	for $x_1 + \frac{\D x_N -x_1}{\D 3} <x_i <x_1 + 2 \frac{\D x_N -x_1}{\D 3}$	\\
		$yMin2$				& $\min\limits_i(y_i)$	---"---	\\
		\hline
	\end{tabular}
\end{table}
\section{Initial parameter estimates for trigonometric functions} 
\begin{figure}
	\centering
\ifTIKZinline
		\input{./TIKZ-Plots/cosineFunction.tex}
\else
	\includegraphics[width=10cm]{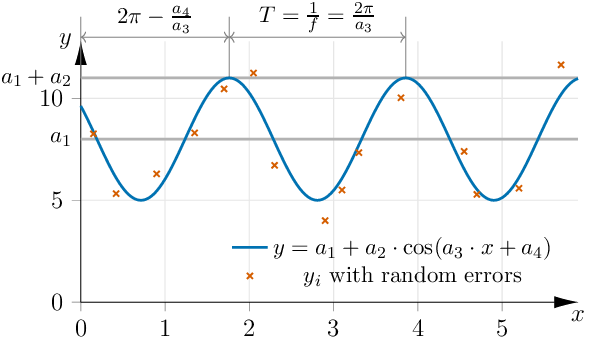}
\fi
	\caption{\label{fig_cosFunc}Example of a trigonometric function and corresponding erroneous observations}
\end{figure}

The trigonometric function (\ref{eq_cosFunc}) describes diverse processes whose appereance depends on four parameters, see \Figu{fig_cosFunc}.
The plot shows erroneous observations $y_i$ and the true trigonometric function.

The following subsections explain how initial estimates of these parameters $a_j$ can be derived from pairs of conditions and observations $(x_i;y_i)$ without specific restrictions on the parameter's domain of definition.
It is solely assumed that (i) the pairs are sorted such that $x_i < x_{i+1}$ holds, and (ii) the conditions can take any value in the interval $[x_1;x_N]$, i.e.\ they do not have to be equidistant.
\subsection{Estimating the offset $a_1$ and the amplitude $a_2$ }
Under the presumption that the conditions $x_i$ are more or less evenly distributed along one ore more periods of the cosine wave, the offset $a_1$ can be simply estimated by averaging all observations $y_i$ and the amplitude $a_2$ can be roughly derived from the range of observations, compare Figure \ref{fig_cosFunc}:
$$
	\hat{a}_1 = \frac{\D 1}{\D N}\cdot \sum_{i=1}^N y_i \qquad \mbox{and} \qquad 
	\hat{a}_2 = 0.5 \cdot \left(\max_i(y_i) - \min_i(y_i) \right)
	\;.
$$
As random noise can increase or decrease the true $y$ values, it is expected that the amplitude $\hat{a}_2$ is systematically estimated too high.
\Algo{alg_preparation} depicts the corresponding pseudo-code. Besides the determination of $\hat{a}_1$ and $\hat{a}_2$, some other values are also prepared for later use.
\begin{algorithm}
	\captionsetup{style=mystyle}
  \caption{\label{alg_preparation}Estimation of the mean value $\hat{a}_1$ and the amplitude $\hat{a}_2$}
	 \hfil\begin{minipage}[t]{0.85\textwidth}
	\footnotesize
	\begin{algorithmic}[1]
		\LComment {input: $N \dots $ number of observations $y_i$, $i=1, 2, \dots, N$}
		\LComment {input: pairs $(x_i,y_i)$, with $x_{i}\le x_{i+1}$}
		\LComment {output: $yMax, yMin \dots $ extremal values of observations}
		\LComment {output: $idxMax, idxMin \dots $ indices of extremal point positions}
		\LComment {output: $yMax2, yMin2 \dots $ extremal values in inner third of condition range}
		\LComment {output: $idxMax2, idxMin2 \dots $ indices of inner extremal point positions}
		\LComment {output: $countMid \dots $ number of inner points}
		\LComment {output: $\hat{a}_1, \hat{a}_2 \dots $ estimated mean and amplitude of cosine function}
		\State {$yMax \gets y_1$; $idxMax \gets 1$} 
		\State {$yMin \gets y_1$; $idxMin \gets 1$} 
		\State {$sumY \gets y_1$} 
		\State {$yMax2 \gets y_{N/2}$; $idxMax2 \gets N/2$} 
		\State {$yMin2 \gets y_{N/2}$; $idxMin2 \gets N/2$} 
		\State {$midLow \gets x_1 + (x_N -x_1)\cdot 0.333$} \Comment{lower bound of inner third}
		\State {$midHigh \gets x_1 + (x_N -x_1)\cdot 0.667$} \Comment{upper bound of inner third}
		\State {$countMid \gets 0$} \Comment{ensure that there are enough observations in the inner third region}
		\For {$i \gets 2; i \le N; i\gets i+1$} \Comment {compare all adjacent observations}
			\State {$sumY \gets sumY + y_i$} \Comment {accumulate observation values}
			\IIf {$yMax < y_i$} {$yMax \gets y_i$; $idxMax \gets i$}
			\EndIIf
			\IIf {$yMin > y_i$}	{$yMin \gets y_i$; $idxMin \gets i$}
			\EndIIf
			\If {$x_i \ge midLow$ AND $x_i \le midHigh$} \Comment {within bounds of inner third}
				\State {$countMid \gets countMid+1$}
				\IIf {$yMax2 < y_i$} {$yMax2 \gets y_i$; $idxMax2 \gets i$}
				\IIf {$yMin2 > y_i$} {$yMin2 \gets y_i$; $idxMin2 \gets i$}
			\EndIf
		\EndFor
		\State {$\hat{a}_1 \gets sumY / N $} \Comment{estimated mean of all observations}
		\State {$\hat{a}_2 \gets 0.5\cdot (yMax - yMin) $}  \Comment{estimated amplitude of cosine function}
	\end{algorithmic}

	\end{minipage}
\end{algorithm}
\subsection{Frequency estimation} 
Estimating the frequency parameter $a_3$ is the most difficult part, as its value is crucial for choosing a suitable position in the error landscape.
If the samples were taken at equal intervals, the frequency could be easily determined using the discrete Fourier transform. For non-uniformly sampled observations, other methods must be considered. A suitable heuristic consists of determining the positions of the zero crossings, or more precisely, the crossings of the mean value $\hat{a}_1$.

Crossings are detected if neighbouring observations $y_i$ and $y_{i-1}$ are located on opposite sides of the estimated mean $\hat{a}_1$, see \Algo{alg_findCrossings}, line \ref{alg2_oppositeSides}.
\begin{algorithm}
	\captionsetup{style=mystyle}
  \caption{\label{alg_findCrossings}Finding positions where the cosine function crosses the mean value $\hat{a}_1$}
	 \hfil\begin{minipage}[t]{0.75\textwidth}
	\footnotesize
	\begin{algorithmic}[1]
		\LComment {input: $N \dots $ number of observations $y_i$, $i=1, 2, \dots, N$}
		\LComment {input: pairs $(x_i,y_i)$, with $x_{i}\le x_{i+1}$}
		\LComment {input: $\hat{a}_1 \dots $ estimated mean of cosine function}
		\LComment {output: $numX \dots $ number of crossings}
		\LComment {output: $crossings[\cdot] \dots $ array of found crossing positions}
		\LComment {output: $meanDev[\cdot] \dots $ array of average deviations from $\hat{a}_1$}
		\State {$numObs \gets 1$} \Comment{initialize counter of observations}
		\State {$sumObs \gets y_1$} \Comment{initialize sum of observations}
		\State {$numX \gets 0$} \Comment{initialize counter}
		\For {$i \gets 2; i \le N; i\gets i+1$} \Comment {compare all adjacent observations}
			\If {$(y_{i-1}>\hat{a}_1)$ AND $(y_i<\hat{a}_1)$ OR 
		       $(y_{i-1}<\hat{a}_1)$ AND $(y_i >\hat{a}_1)$} 	\label{alg2_oppositeSides}
				\LComment {crossing found}
				\LComment {perform straight-line approximation $y=m\cdot x+n$}	\label{alg2_straightLineStart}
				\If {$(x_i - x_{i-1}) > 10^{-8}$} \Comment{avoid numerical instability}
					\State {
							$$
							m \gets \frac{y_i - y_{i-1}}{x_i - x_{i-1}} \qquad
							n \gets y_i - m\cdot x_i
							$$
							}
					\LComment {store estimated mean-crossing position $x_{\rm c}$ based on 
								$\hat{a}_1 = m\cdot x_{\rm c} + n$}
					\State {$crossings[numX] \gets (\hat{a}_1 - n) / m$}		\label{alg2_straightLineEnd}
				\Else
					\State {$crossings[numX] \gets x_i$}
				\EndIf
				\LComment{save average absolute difference to $\hat{a}_1$ within segment}
				\State $meanDev[numX] \gets |sumObs / numObs - \hat{a}_1|$
				\State {$numX \gets numX + 1$}
				\State {$sumObs \gets y_i$; $numObs \gets 1$} \Comment{start with next segment}
			\Else
				\State {$sumObs \gets sumObs + y_i$; $numObs \gets numObs+ 1$} \Comment{accumulate values}
			\EndIf
		\EndFor
	\end{algorithmic}

	\end{minipage}
\end{algorithm}
Using a straight-line approximation between two observations, the intersection position can be roughly determined, lines \ref{alg2_straightLineStart}-\ref{alg2_straightLineEnd}. Naturally, these computations are disturbed by the errors in $y_i$ and $y_{i-1}$.
All crossing positions are stored for further inspection.
The main idea behind this approach is to use a representative distance between these crossing as the half of one oscillation period $T=1/f$.

Depending on the position of the samples, sample density, and observation errors, it is likely that intersection points will be recorded that are closer together or further apart than the theoretical ones.  
If there are only few samples per oscillation, as in Figure \ref{fig_cosFunc}, we can assume that distances similar to the correct one build the majority of distances.
After sorting all measured distances in ascending order, the distance in the middle (the median) would be a good representative of $T/2$.

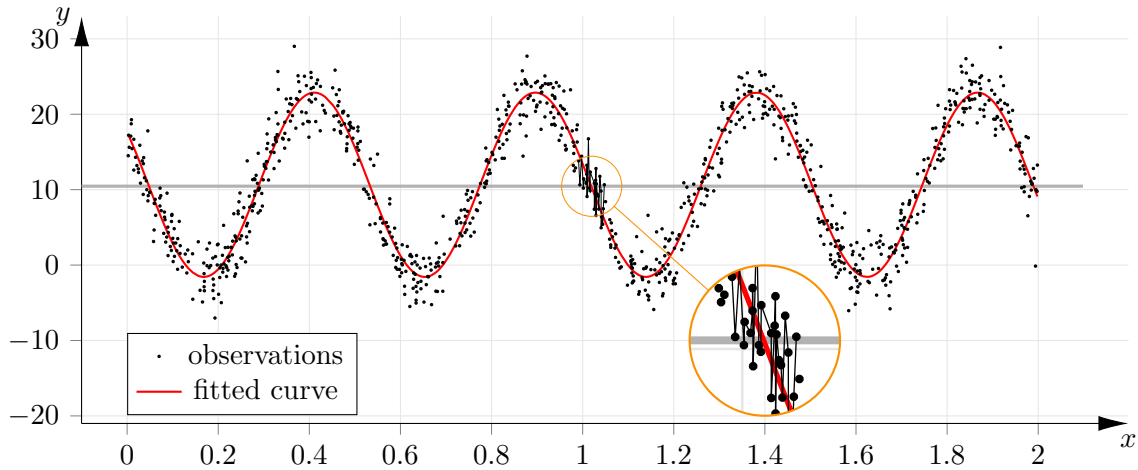
\begin{figure}
	\centering
	\pgfmathsetmacro{\iaone}{10.468}   %
\pgfmathsetmacro{\iatwo}{18.01}  	 %
\pgfmathsetmacro{\iathree}{132.49} %
\pgfmathsetmacro{\iafour}{1.632}   %

\pgfmathsetmacro{\aone}{10.46}   %
\pgfmathsetmacro{\atwo}{0.8156}   %
\pgfmathsetmacro{\athree}{134.2} %
\pgfmathsetmacro{\afour}{0.6370}   %

\pgfplotstableread[header=false]{%
Example/cosine_samples1000.txt}\mytable

\newcommand{\findminmax}[1]{%
  \pgfplotstablegetrowsof{\mytable}
  \pgfmathtruncatemacro{\numrows}{\pgfplotsretval-1}
  \typeout{\numrows\space rows}
  \pgfplotstablegetelem{0}{#1}\of{\mytable}
  \pgfmathtruncatemacro{\mymax}{\pgfplotsretval}
  \pgfmathtruncatemacro{\mymin}{\pgfplotsretval}
  \typeout{initially:\space\mymin}
  \pgfplotsinvokeforeach {1,...,\numrows}{
    \pgfplotstablegetelem{##1}{#1}\of{\mytable}
    \pgfmathsetmacro{\mymax}{max(\pgfplotsretval,\mymax)}
    \pgfmathsetmacro{\mymin}{min(\pgfplotsretval,\mymin)}
   }
}
\findminmax{0}
\let\xmaxinit=\mymax
\let\xmininit=\mymin
\findminmax{1}
\let\ymaxinit=\mymax
\let\ymininit=\mymin

\def\xmin{\xmininit-0.1}
\def\xmax{\xmaxinit+0.1}
\def\ymin{-21}
\def\ymax{33}

\begin{tikzpicture}[scale=1,
  spy using outlines={circle, thick, magnification=2.5, size=2.cm, connect spies}]
	
  \begin{axis}[
    width=15.5cm, height=7cm,
    xmin=\xmin, xmax=\xmax+0.1,
    ymin=\ymin, ymax=\ymax,
    axis lines=left,
		axis line style={-{Latex[length=4mm,width=2.mm]}},
		xlabel style = {at={(axis description cs:1.,0.0)},anchor=north,font=\small},
		ylabel style = {at={(axis description cs:0.0,1)},anchor=east,rotate=-90,font=\small},
    xlabel={$x$}, ylabel={$y$},
    tick align=outside,
		ticklabel style={font=\small},
    grid=both,
    major grid style={gray!20},
    minor grid style={gray!10},
    legend style={font=\small,at={(axis cs:0.0,-9)},anchor=north west},
    clip=false,
		restrict y to domain=-10.1:30.1 %
  ]

    \addplot[
			forget plot, %
			domain=\xmin:\xmax, samples=2, very thick, color=black!30]
      {\aone};

    \addplot[
      thick,
      color=black,
			mark=*, mark size=0.25pt,
			only marks
    ] table[x index=0, y index=1] {\mytable};
		\addlegendentry{\small observations}

    \addplot[
      thick,
      color=red,
    ] table[x index=0, y index=2] {Example/example_result1000.txt}; %
		\addlegendentry{\small fitted curve}

    \addplot[
			forget plot, %
      very thin,
      color=black,
			restrict x to domain=0.99:1.05,
    ] table[x index=0, y index=1] {\mytable};

		\coordinate (A) at (1.02,10.46);
		\coordinate (B) at (1.4,-10);
		\spy [myorange] on (A) in node at (B);

  \end{axis}
	
\end{tikzpicture}
\ifTIKZinline
\else %
\fi
	\caption{\label{fig_cosFuncNoisy}Example of a trigonometric function with dense sampling and strong noise. The magnified section shows how the noisy curve crosses the centre line $a_1\approx 10.5$ several times (spurious crossings).}
\end{figure}

In scenarios where a high number of (erroneous) samples are recorded per oscillation, there is a relevant number of spurious crossings that are very close in time and occur around the theoretical crossings \cite{Gri12}, see \Figu{fig_cosFuncNoisy}. 
This number may be higher than the number of correct crossing distances. In this case, the median distance would not lead to a suitable result. This problem could be solved by additional processing methods such as pre-filtering or piece-wise straight-line approximations, see for instance \cite{Men14}.
The next subsections propose a simpler approach.
\subsubsection{Analysis and presumptions regarding the signal properties} 
\begin{figure}
	\centering
\ifTIKZinline
	\input{./TIKZ-Plots/cos_noiseCxx_N0100.tex}
\else
	\includegraphics[width=15.5cm]{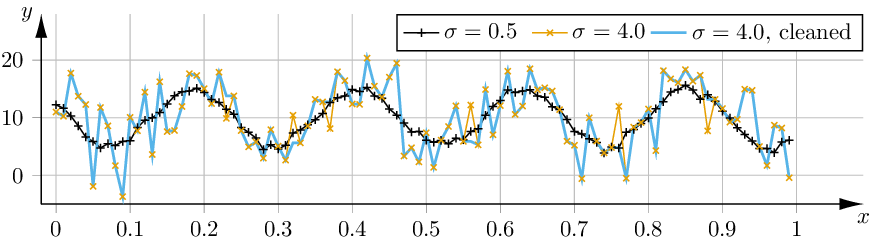}
\fi
	\caption{\label{fig_cos_noiseC_N0100_plt}Example of a trigonometric function with two different noise levels $\sigma$}
\end{figure}

\begin{figure}
	\begin{minipage}{0.499\textwidth}\centering
	\includegraphics[width=\textwidth]{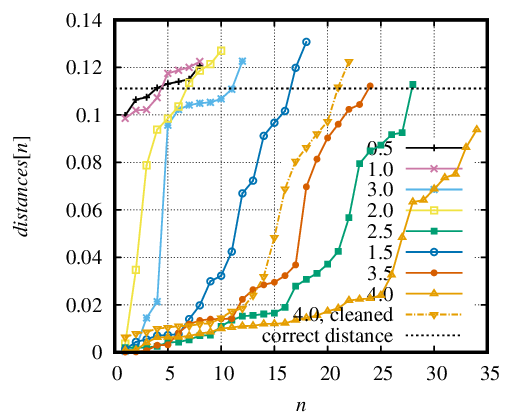} \\
	(a)
	\end{minipage}
	\hfil
	\begin{minipage}{0.499\textwidth}\centering
	\includegraphics[width=\textwidth]{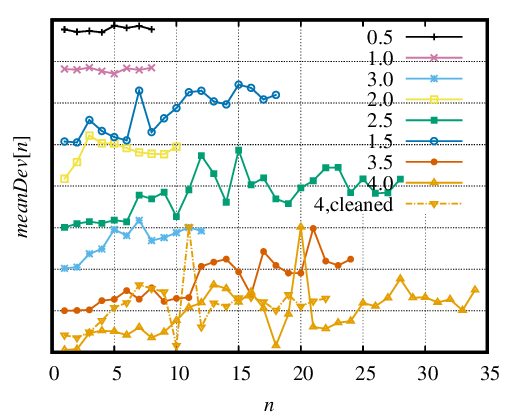} \\
	(b)
	\end{minipage}
	\caption{\label{fig_distancesN0100}Sorted values dependent on the noise level (standard deviation): (a) distances between crossings, (b) corresponding averaged absolute deviation of observation values $y_i$ from estimated mean $\hat{a}_1$}
\end{figure}

A first analysis has been done using a trigonometric function with arbitrarily chosen parameters $a_1=10$, $a_2=5$, $f=4.5$ ($a_3=2\pi\cdot 4.5$), and $a_4 = \pi/3$. It has been sampled with a sampling frequency of $f_{\rm s}=100$ Hz and noise drawn from a normal distribution ${\cal N}(\mu=0, \sigma^2)$ has been added. The strength of the noise has gradually been increased from $\sigma=0.5$ to  $\sigma=4$, see \Figu{fig_cos_noiseC_N0100_plt}.
The distances between the crossing positions have been measured and sorted in ascending order. \Figu{fig_distancesN0100}(a) visualizes these distances for different noise levels. 
Corresponding to the chosen frequency, the expected distance is $d_{true} = 0.5\cdot T = 0.5/f = 0.5/4.5=0.\ol{11}$.
It is marked with a horizontal line.

It can be seen in Figure \ref{fig_distancesN0100}(a) that at the two lowest noise levels, nine crossings could be found, resulting in eight distances that are closely distributed around the correct distance. At higher noise levels, false crossings occur, leading to relatively short distances. The resulting number of (incorrect) crossings correlates only roughly with the noise level. For example, noise with $\sigma=2.5$ in this example test causes more crossings by chance than $\sigma=3.5$.
For low noise conditions, there is a distinct gap between between short (and false) distances and long distances, which has been utilized for example in \cite{Ang14}. However, with increasing noise, both types of distances can be come very close and other considerations are required. 

\Figu{fig_distancesN0100}(b) shows additional information about the average deviation of signal values $y_i$ from estimated mean value of all observations for each segment:
		\begin{align}
			meanDev[n] = \frac{1}{N_S}\sum_{i\in S_n}|y_i -\hat{a}_1| \qquad 
			\begin{array}{l}
				S \dots \mbox{segment between two crossings}  \\
			  N_S \dots \mbox{number of observations in segment~} S
			\end{array}
			\;.
		\end{align}
It had been assumed that the observations between correct crossings cover a more or less full half-wave of the cosine function and the absolute deviation should be high. False crossings are mostly located where the cosine intersects its mean and the deviation from the mean should be low. In general, the plots in Figure \ref{fig_distancesN0100}(b) reflect this trend. In absence of false crossings ($\sigma \le 1.0$) this deviations is almost constant. For $1.0 < \sigma \le 2.0$, the values of the eight longest (and good) distances are higher than the values of the shorter distances. The higher the noise, the less this behaviour is pronounced ant these deviation values could not been utilized. 

Based on the analysis, following presumptions can be made:
\begin{enumerate}[label=/\arabic*/]
	\item \label{pre_goodVSspurious} Regardless of the noise level, an incorrect (spurious) distance $d_{\rm s}$ cannot be greater than half the correct distance $d_0$.
	\item \label{pre_spreadOfSpurious} Since false crossings occur randomly, we can assume that the corresponding distances $d_{{\rm s},j}$ vary significantly:
		\begin{align}\label{eq_spuriousDistCondition}
			\max_j(d_{{\rm s},j}) > 2\cdot \min_j(d_{{\rm s},j})
			\;.
		\end{align}
	\item \label{pre_noSpurious}Complementary to \ref{pre_spreadOfSpurious} it can be stated that the shortest estimated good distance $\hat{d}_0$ is not expected to be shorter than 50\% of the longest good distance:
		\begin{align}\label{eq_goodDistCondition}
			0.5\cdot \max_i(\hat{d}_{0,i}) \le \min_i(\hat{d}_{0,i})
			\;.
		\end{align}
	\item \label{pre_uncertainRange}Distances close to the half of the correct distance $d_{0}$ are highly uncertain in terms of their correct class (spurious vs. good). So, a suitable threshold must be applied.
	\item \label{pre_spuriousPairs}Spurious crossings always appear in pairs. Therefore, the number of false distances is even.
	\item \label{pre_sumOfSpuriousDistances}The more spurious crossings occur, the shorter the observed `good' distances are. This is simply because the sum of all distances is nearly constant.\footnote{Boundary effects may cause a slightly varying total distance between the outer crossings.} Therefore, after separating both clusters, a correction term should be added to the estimated distance parameter.
	\item \label{pre_sumOfDistances}The sum of all distances is higher than one third of the entire range of conditions (noise-free case):
		\begin{align}
			\sum_{k} d_k >  \frac{x_N - x_1}{3}
			\;.
		\end{align}
		The minimum case occurs when the sampled signal contains three half-waves with only one half-wave in the middle enclosed by two crossings.
	\item \label{pre_extremalDistance}The distance between the extremal point positions is roughly an odd multiple of the correct distance:
		\begin{align}
			imax=\argmax_i(y_i)&\qquad imin=\argmin_i(y_i) \nonumber\\
			|x_{imax} - x_{imin} | &\approx m \cdot d_0 \qquad m=1,3,5,\dots
			\;.
		\end{align}
	\item \label{pre_spikes}Single signal values, which lie on the opposite side of $a[0]$ compared to their direct neighbours, can split good distances into unfavourable parts. If they have a smaller magnitude than their neighbours (with respect to $a[0]$), they can be classified as disturbing spikes and should be removed.
\end{enumerate}

\subsubsection{Proposed determination of the frequency parameter $a_3$} 
Based on the presumptions from the previous Subsection, an algorithm can be developed which robustly determines a suitable estimate $\hat{a}_3$.

Extensive investigations have shown that the processing of distances between crossings can be significantly improved by accounting for presumption \ref{pre_spikes}. Especially, in the presences of strong noise in the signal, these spikes disturb the measurement of good distance, and removing these spikes is a crucial pre-processing step, see \Algo{alg_removeSpikes}.
\begin{algorithm}
	\captionsetup{style=mystyle}
  \caption{\label{alg_removeSpikes}Pre-filtering the vector of observations by spike removal}
	 \hfil\begin{minipage}[t]{0.85\textwidth}
	\footnotesize
	\begin{algorithmic}[1]
		\LComment {input: ${\mathbf y} \dots $ vector of observations}
		\LComment {input: $N \dots $ number of observations $y_i$, $i=1, 2, \dots, N$}
		\LComment {input: $\hat{a}_1 \dots $ estimated mean of cosine function}
		\LComment {output: $\mathbf{y_{\rm c}} \dots $ vector of cleaned observations}
		\State{$y_{{\rm c},1} \gets y_1$} \Comment{copy first value}
		\For {$i \gets 2; i \le N-1; i\gets i+1$} \Comment {compare triplets of observations}
			\If {$(y_{i-1}>\hat{a}_1)$ AND $(y_i<\hat{a}_1)$ AND $(y_{i+1}>\hat{a}_1)$ }
					\Comment {negative spike found}
				\State {$minY \gets \min(y_{i-1},y_{i+1})$}
				\If {$minY - \hat{a}_1 > \hat{a}_1 - y_i$}
					\Comment{$|y_{i}-\hat{a}_1| < |y_{i-1}-\hat{a}_1|, |y_{i+1}-\hat{a}_1|$}
					\LComment {absolute spike is smaller than neighbours}
					\State {$y_{{\rm c},i} \gets minY$}
						\Comment {set to the closest positive neighbour}
				\Else
					\State{$y_{{\rm c},i} \gets y_i$} \Comment{copy original value}
				\EndIf
			\ElsIf {$(y_{i-1}<\hat{a}_1)$ AND $(y_i>\hat{a}_1)$ AND $(y_{i+1}<\hat{a}_1)$ }
					\Comment {positive spike found}
				\State {$maxY \gets \max(y_{i-1},y_{i+1})$}
				\If {$\hat{a}_1 - maxY >  y_i  -\hat{a}_1$}
					\Comment {absolute spike is smaller than neighbours}
					\State {$y_{{\rm c},i} \gets maxY$}
						\Comment {set to the closest negative neighbour}
				\Else
					\State{$y_{{\rm c},i} \gets y_i$} \Comment{copy original value}
				\EndIf
			\Else
					\State{$y_{{\rm c},i} \gets y_i$} \Comment{copy original value}
			\EndIf
		\EndFor
		\State{$y_{{\rm c},N} \gets y_N$} \Comment{copy last value}
	\end{algorithmic}

	\end{minipage}
\end{algorithm}
The spike removal decreases the number of crossings and it becomes more unlikely that a good distance is split into large spurious distances. An example is shown in Figure \ref{fig_cos_noiseC_N0100_plt}. The plot entitled with '$\sigma=4.0$, cleaned` shows the curve after the removal of six spikes. This causes a reduction of the number of distances from $n=34$ to 22 number, see Figure \ref{fig_distancesN0100}.
Based on the cleaned observations vector $\mathbf{y}_{\rm c}$, the search for the correct crossing distances can be started.\footnote{Note that already Algorithm \ref{alg_findCrossings} must be performed based on $\mathbf{y}_{\rm c}$ instead of $\mathbf{y}$.}

Logically, a distance can only be computed if there are at least two crossings, \Algo{alg_getBestDistance}, lines 5-12.
\begin{algorithm}
	\captionsetup{style=mystyle}
  \caption{\label{alg_getBestDistance}Select representative crossing distance for frequency estimation}
	  \hfil\begin{minipage}[t]{0.87\textwidth}
	\footnotesize
	\def\rI{re\!f\!Idx}

\begin{algorithmic}[1]
\LComment {input: $numX \dots $ number of crossings}
\LComment {input: $idxMax, idxMin \dots $ indices of extremal point positions}
\LComment {input: $crossings[\cdot] \dots $ array of found crossing positions}
\LComment {output: $\hat{a}_3 \dots $ angular frequency}
\If {$numX > 1$}
	\State{$numDists = numX -1$} \Comment{number of distances}
	\State{$sumO\!f\!Distances \gets 0$}
	\For {$i \gets 1; i \le numDists; i\gets i+1$} \Comment {all crossings; leave index 0 unused}
			\LComment {compute distance between crossings}
			\State {$distances[i] \gets crossings[i+1] - crossings[i]$}
			\State {$sumO\!f\!Distances \gets sumO\!f\!Distances+distances[i]$}
				\Comment{for sanity check}
	\EndFor
	\If {$numDists > 1$ AND $sumO\!f\!Distances < (x_N - x_i) / 3$}
		\State {$numX \gets 1$} \Comment{something is wrong, assume fraction of a cosine}
	\Else
		\State {sort values in $distances[\cdot]$ in ascending order; sort $meanDev[\cdot]$ accordingly}
		\LComment{select representative distance}
		\If {$distances[1] \cdot 2 > distances[numDists]$}
			\Comment{no outliers}
			\State {$d^* \gets sumO\!f\!Distances/ numDists$} 		\Comment{take average distance}
		\Else \Comment{expect spurious distances; do classification}\label{alg4_3stepProcStart}
			\State {$d_{\rm ref}, \rI \gets$ {\scshape getReferenceDistance($numDists$, $distances[\cdot]$)}}
				\Comment {function call}
			\State {$d_{\rm typ}, numO\!f\!GoodDists \gets$ {\scshape getTypicalDistance( $d_{\rm ref}$, $\rI$, $numDists$ \dots)}}
			\LComment{add spurious distances to good distances}
			\State {$sumO\!f\!SpuriousDists \gets 0.0$};
			\For {$i \gets numDists-numO\!f\!GoodDists; i > 0; i\gets i-1$} \Comment {summation of spurious distances}
				\State {$sumO\!f\!SpuriousDists \gets sumO\!f\!SpuriousDists + distances[i]$};
			\EndFor
			\State{$correctionTerm \gets \frac{sumO\!f\!SpuriousDists}{numO\!f\!GoodDists}$} \label{alg4_correctionStart}
			\State{$d^* \gets d_{\rm typ} + \min(correctionTerm, d_{\rm typ})$}\Comment{max. change 100\%}\label{alg4_correctionEnd}
		\EndIf  \label{alg4_3stepProcEnd}
		\State{$\hat{a}_3 \gets \pi / d^*$}
			\Comment{$\hat{a}_3= 2\pi\cdot f = 2\pi/T = \pi / (T\cdot 0.5)$}
	\EndIf
\EndIf
\If {$numX \le 1$}
		\Comment{presume that conditions cover a half oscillation}
	\State {$\hat{a}_3 \gets \pi \cdot(x_N - x_i)$}
\EndIf

\end{algorithmic}

	\end{minipage}
\end{algorithm}
At the same time, the sum of all distances found is calculated for verification. According to \ref{pre_sumOfDistances}, the sum has to exceed a minimum. If it is to low, then one must assume that only a fraction of a cosine wave is contained and there is merely a single true crossing, lines 13-14.
When every thing seems to be reasonable, the distances are sorted in ascending order and the absence of spurious distances is identified by comparing the first (i.e. the shortest) and the last (i.e. the longest) distance. If presumption \ref{pre_noSpurious} holds ($distances[0] \cdot 2 > distances[numDists-1]$), then there are no false distances and taking the average distance as a representative of $T/2$ is possible, lines 18-19:
	\begin{align}
		d^* = \frac{sumO\!f\!Distances}{numDists}
		\;.
	\end{align}
Otherwise, a tree-step method is applied, lines \ref{alg4_3stepProcStart}-\ref{alg4_3stepProcEnd}:
\begin{enumerate}
	\item determination of a reference distance $d_{\rm ref}$ that allows the rough discrimination of good and spurious distances,
	\item derivation of a typical distance $d_{\rm typ}$ from the refined class of good ones, and
	\item correction of the typical distance by taking the contribution of spurious distances into account, presumption \ref{pre_sumOfSpuriousDistances}.
\end{enumerate}
\paragraph{Determination of a reference distance:} 
It proved to be sensible to define a threshold that separates spurious distances from the good ones. The tricky part is to find a reference distance to compare with. The problem is that also distances much longer than typical good ones can appear caused by missing half-waves. It is therefore to be expected that there is at least one good distance, while the spurious and longer distances may or may not be present.

The proposed idea is to create a histogram with equally spaced bins and to check whether \ref{pre_noSpurious} is fulfilled within the bins, \Algo{alg_findReferenceDistance}.
\begin{algorithm}
	\captionsetup{style=mystyle}
  \caption{\label{alg_findReferenceDistance}Determination of a reference distance for classification of distances --- Part 1}
	  \hfil\begin{minipage}[t]{0.9\textwidth}
	\footnotesize
\def\rI{re\!f\!Idx}
\def\nBins{numO\!f\!Bins}
\def\rBIdxOne{binIdx1}
\def\rBIdxTwo{binIdx2}
\def\rFIdx{f\!irstIdx}
\def\rLIdx{lastIdx}

	\begin{algorithmic}[1]
		\Function {getReferenceDistance}{$distances[\cdot]$, $numDists$}
		\LComment {input: $distances[\cdot] \dots $ vector of distances; $numDists \dots $ number of distances}
		\LComment {output: $d_{\rm ref} \dots $ reference distance for discrimination between good and spurious}
		\LComment {output: $\rI \dots $ corresponding index}
		\State{$\rBIdxOne \gets -1$} \Comment{no good bins found yet}
		\State{$\nBins \gets 2$} \Comment{start with two bins}
		\State{$stop \gets false$} \Comment{reset flag}
		\Repeat \label{alg5_loopStart}
			\For {$i \gets 0; i < \nBins; i\gets i+1$} \label{alg5_binsStart}
				\State{$binCount[i] \gets 0$; $\rFIdx[i] \gets 0$; $\rLIdx[i] \gets 0$}\Comment{reset values}
			\EndFor
			\State{$binWidth \gets distances[numDists] / \nBins$}
			\State{$th \gets 1.001 \cdot binWidth$}	\Comment{one per mille more ensures that last bin includes last distance}
			\State{$binIdx \gets 0$} \Comment{start collecting of distances for first bin}
			\State{$\rFIdx[binIdx] \gets 1$} \Comment{index = 0 is unused}
			\For {$i \gets 1; i \le numDists; i\gets i+1$} 
				\If {$distances[i] > th$}
					\State {$th \gets th + binWidth$} \Comment{threshold for next bin}
					\State {$binIdx \gets binIdx + 1$} \Comment{next bin}
					\State {$\rFIdx[binIdx] \gets i$} \Comment{remember first index in this bin}
				\EndIf
				\State {$binCount[binIdx] \gets binCount[binIdx] + 1$} \Comment{increment count}
				\State {$\rLIdx[binIdx] \gets i$} \Comment{remember last index in this bin}
			\EndFor \label{alg5_binsEnd}
			\State{$numCandidateBins \gets 0$} \Comment{reset}			\label{alg5_binContiguousStart}
			\For {$i \gets 0; i \le binIdx; i\gets i+1$} \Comment{for all used bins}
				\If{$distances[\rFIdx[i]] \cdot 2 > distances[\rLIdx[i]$}
					\State{$numCandidateBins \gets numCandidateBins + 1$} \Comment{good bin found}
					\IIf {$\rBIdxOne < 0$} {$\rBIdxOne \gets i $} \EndIIf \Comment{remember first bin index}
					\State{$\rBIdxTwo \gets i$} \Comment{update second bin index}
				\EndIf
			\EndFor \label{alg5_binContiguousEnd}
			\State {$\nBins \gets \nBins + 1$} \Comment{increment number of bins}
			\IIf{$numCandidateBins > 1$} {~$stop \gets true$} \Comment{enough good bins found}
			\IIf {$\nBins > 5$} {~$stop \gets true$}  \EndIIf\Comment{solely a single good bin has been found}
		\Until {$stop == true$}  \label{alg5_loopEnd}
		
		~~~~:\\
		{continued in Algorithm \ref{alg_findReferenceDistancePart2}}
		\algstore{getReferenceDistance}
\end{algorithmic}

	\end{minipage}
\end{algorithm}
The method starts with two bins accounting for two different possible settings: (i) the lower bin comprises a bunch of spurious distances and the upper some good ones or (ii) the lower bin contains several good distances and the upper contains one or more long distances caused by missed half-waves.

First, the algorithm counts how many distances fall into the different bins, lines \ref{alg5_binsStart}-\ref{alg5_binsEnd}. For later purpose, the indices of the first and the last distance are stored for each bin.
Afterwards, for each bin it is checked whether \ref{pre_noSpurious} holds, lines \ref{alg5_binContiguousStart}-\ref{alg5_binContiguousEnd}.
If the answer is yes, the corresponding bin indices are saved.
As soon as two possibly spurious-free bins have been found, the reference distance can be derived. Otherwise, the number of bins is incremented and the procedure goes into iteration until either two matching bins can be found or a maximum number of five bins has been tested, \ref{alg5_loopStart}-\ref{alg5_loopEnd}. 

Depending on the stopping reason, different decisions must be made, \Algo{alg_findReferenceDistancePart2}.
\begin{algorithm}
	\captionsetup{style=mystyle}
  \caption{\label{alg_findReferenceDistancePart2}Determination of a reference distance for classification of distances --- Part 2}
	  \hfil\begin{minipage}[t]{0.9\textwidth}
	\footnotesize

\def\rI{re\!f\!Idx}
\def\nBins{numO\!f\!Bins}
\def\rBIdxOne{binIdx1}
\def\rBIdxTwo{binIdx2}
\def\rFIdx{f\!irstIdx}
\def\rLIdx{lastIdx}

	\begin{algorithmic}[1]
		\algrestore{getReferenceDistance}
		\item{~~~~:} %
		\If {$\nBins \le 5$}\Comment{two good bins have been found}
			\State{$num \gets \rLIdx[\rBIdxTwo] - \rFIdx[\rBIdxOne] + 1$}
			\State{$\rI \gets \rFIdx[\rBIdxOne] + (num >> 1)$}
							 \Comment{median distance from both bins}
			\If {$\rBIdxTwo == \rBIdxOne +1$} \label{alg6_adjacentBins}\Comment{bins are neighbours}
				\If {$distances[\rFIdx[\rBIdxOne]] \cdot 2 > distances[\rLIdx[\rBIdxTwo]]$}\label{alg6_ifMergedBins}
					\LComment{combination of bins also fulfills criterion, compute the average of distances}
					\State{$sum \gets distances[\rFIdx[\rBIdxOne]]$} \label{alg6_calcAverageOfDistsStart} \Comment{initialize sum of distances}
					\For {$i \gets \rFIdx[\rBIdxOne]]+1; i \le \rLIdx[\rBIdxTwo]; i\gets i+1$} 
						\State{$sum \gets sum + distances[i]$}\Comment{summation of all distances in these two bins}
					\EndFor
					\State{$d_{\rm ref} \gets \frac{\D sum }{\scriptstyle\rLIdx[\rBIdxTwo] - \rFIdx[\rBIdxOne] + 1}$}
							 \label{alg6_calcAverageOfDistsEnd} \Comment{average distance}
				\Else
					\State{$d_{\rm ref} \gets distances[\rI]$}\label{alg6_takeMedianOfDists}
				\EndIf
			\Else
				\State{$num \gets \rLIdx[\rBIdxTwo] - \rFIdx[\rBIdxTwo]$}\label{alg6_calcMedianHighBinStart} 
				\State{$\rI \gets \rFIdx[\rBIdxTwo] + (num >> 1)$}
				\State{$d_{\rm ref} \gets distances[\rI]$}
						 \Comment{median distance from last bin}\label{alg6_calcMedianHighBinEnd} 
			\EndIf
		\Else \Comment{process single bin}\label{alg6_singleBinStart} 
			\State{$\rI \gets numDists$} \Comment{there is a single good distance}
			\State{$d_{\rm ref} \gets distances[\rI]$}		\label{alg6_singleBinEnd} 
		\EndIf
		\State \Return $d_{\rm ref}$, $\rI$
			\Comment{return selected reference distance }
	\EndFunction
\end{algorithmic}

	\end{minipage}
\end{algorithm}
If the number of bins remained in the allowed limit, two bins have been found likely containing good distances. If they are neighbours, line \ref{alg6_adjacentBins}, it is tested whether \ref{pre_noSpurious} is still valid after merging the two bins, line \ref{alg6_ifMergedBins}. In lines \ref{alg6_calcAverageOfDistsStart}-\ref{alg6_calcAverageOfDistsEnd}, the average of all distances of the two bins is computed and taken as reference distance. In case that the distances of the two bins vary too much, the median of all distances is the better choice, line \ref{alg6_takeMedianOfDists}. If the two bins are not neighbours, the median of distances from the bin containing the longer distances is taken as reference distance\footnote{This behaviour has not yet been observed and is implemented as fallback routine.}, lines \ref{alg6_calcMedianHighBinStart}-\ref{alg6_calcMedianHighBinEnd}.

At last we have to discuss the case, when only a single bin satisfying \ref{pre_noSpurious} could be found, lines \ref{alg6_singleBinStart}-\ref{alg6_singleBinEnd}. 
This case can occur when there is only a single good distance and some very short spurious distances and we can immediately redirect $d_{\rm ref}$ to the maximum distance.
\paragraph{Derivation of the typical distance:} 
\Algo{alg_findTypicalDistance} shows how a typical distance $d_{\rm typ}$ is derived from the collected information.
\begin{algorithm}
	\captionsetup{style=mystyle}
  \caption{\label{alg_findTypicalDistance}Derivation of a typical distance from the class of good distances}
	  \hfil\begin{minipage}[t]{0.9\textwidth}
	\footnotesize
\def\rI{re\!f\!Idx}

	\begin{algorithmic}[1]
		\Function {getTypicalDistance}{$distances[\cdot]$, $numDists$, $limitDistance$}
		\LComment {input: $distances[\cdot], numDists \dots $ vector of distances and number of distances}
		\LComment {input: $d_{\rm ref}, \rI \dots $ reference distance and its index}
		\LComment {output: $d_{\rm typ} \dots $ typical distance from class of good distances}
		\LComment {output: $numO\!f\!GoodDists   \dots $ number of good distances}
		\LComment {output: $longDistanceFlag \dots $ number of spurious distances}
		\State{$f\!actorLow \gets 2.0$, $f\!actorHi \gets 3.1$}\Comment{compute limits relative to $d_{\rm ref}$}
		\LComment{sanity check for $d_{\rm ref}$} \label{alg7_sanityCheckStart}
		\If {$\rI+1 == numDists$} \Comment{$d_{\rm ref}$ is located at second highest distance}
			\If {$d_{\rm ref} \cdot f\!actorHi < distances[numDists]$} 
				\State{} \Comment{$d_{\rm ref}$ is significantly lower than maximum distance}
				\State{$\rI \gets numDists$} \Comment{assume that highest distance is the only good distance}
				\State{$d_{\rm ref} \gets distances[numDists]$}
			\EndIf
		\EndIf \label{alg7_sanityCheckEnd}
		\LComment{check presence of long distance at index $=numDists$}
		\State{$longDistanceFlag \gets false$}\label{alg7_longDistStart}
		\If{$distances[numDists] > d_{\rm ref} \cdot f\!actorLow$ AND 
		    $distances[numDists] < d_{\rm ref} \cdot f\!actorHi$}
			\State{$longDistanceFlag \gets true$}\Comment{max. distance lies in a typical range}
		\EndIf		\label{alg7_longDistEnd}
		\For {$i \gets \rI-1; i \ge 0; i\gets i-1$} \label{alg7_determineGoodIdxStart}
		\Comment {find border below reference index}
			\If {$distances[i] \cdot 2.3 < d_{\rm ref}$}
				\State {$goodIdx \gets i + 1$, break}  \Comment{distance is too low to be good, leave for-loop}
			\EndIf
		\EndFor		\label{alg7_determineGoodIdxEnd}
		\State {$numO\!f\!GoodDists \gets numDists - goodIdx + 1$}
		\LComment{select median of good distances}
		\If {$\mbox{modulo}(numO\!f\!GoodDists, 2) == 1$} \label{alg7_determineMedianStart}
			\State {$d_{\rm typ} \gets distances[goodIdx + (numO\!f\!GoodDists >> 1)]$} 
		\Else
			\State {$d_{\rm typ} \gets 0.5\cdot (distances[goodIdx + (numO\!f\!GoodDists >> 1)] ~+$}
			{\phantom{$d_{\rm typ} \gets 0.5\cdot (W$}$distances[goodIdx + (numO\!f\!GoodDists >> 1) - 1])$}
		\EndIf		\label{alg7_determineMedianEnd}
		\State{$sum \gets 0.0$}%
									\Comment{additionally compute the mean of good distances}
		\For {$i \gets goodIdx; i \le numDists; i\gets i+1$}
			\State{$sum \gets sum + distances[i]$}%
		\EndFor
		\IIf {$longDistanceFlag == true$} \label{alg7_determineMeanStart}
			{$numO\!f\!GoodDists \gets numO\!f\!GoodDists + 1$} 
		\EndIIf
		\State {$d_{\rm typ} \gets 0.5\cdot (d_{\rm typ} + sum / numO\!f\!GoodDists)$}\Comment{average median and mean value}
		\State \Return $d_{\rm typ}$, $numO\!f\!GoodDists$		\label{alg7_determineMeanEnd}
	\EndFunction
\end{algorithmic}

	\end{minipage}
\end{algorithm}

First, a sanity check is performed. When the $d_{\rm ref}$ points to the second highest distance and it is distinctly lower than the maximum distance, then the maximum distance is most likely the only good distance. The reference distance is change accordingly, lines \ref{alg7_sanityCheckStart}-\ref{alg7_sanityCheckEnd}.
All factors used in this algorithm are based on empirical studies.

Second, a long distance is detected if it is (i) at least twice as long as $d_{\rm ref}$ and (ii) it is not longer than 3.1 times the $d_{\rm ref}$, lines \ref{alg7_longDistStart}-\ref{alg7_longDistEnd}.
The latter condition takes into account the case where the reference distance is incorrectly within the region of spurious distances, for example, if there is only one good distance.
The resulting flag is later used for correcting the number of good distances.

\begin{figure}
	\centering
\ifTIKZinline
	\input{./TIKZ-Plots/cosineFractions.tex}
\else
	\includegraphics[width=10cm]{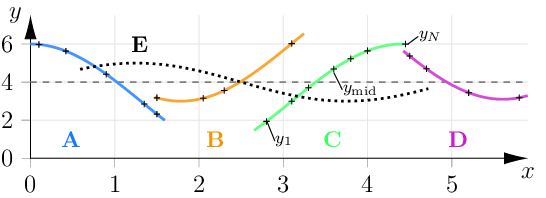}
\fi
	\caption{\label{fig_cosFrac}Examples of fractions of a cosine function with a single crossing of the mean and different types of curvature}
\end{figure}

The reference distance $d_{\rm ref}$ times 2.3 serves as threshold for the discrimination of good and putative spurious distances, lines \ref{alg7_determineGoodIdxStart}-\ref{alg7_determineGoodIdxEnd}.  
The rationale is that (i) spurious distances are the short ones according to \ref{pre_goodVSspurious} and they should be excluded from computations, presumption \ref{pre_uncertainRange}. 
The median position within the range of good distances is taken, lines \ref{alg7_determineMedianStart}-\ref{alg7_determineMedianEnd}:
$$
	d_{\rm typ} \gets \median_i(distances[i] \;|\;goodIdx \le i \le numDists)
	\;.
$$
The result is then combined with the mean value of all good distances, lines \ref{alg7_determineMeanStart}-\ref{alg7_determineMeanEnd}. This averages out random errors in both measurements and can improve the result.
\paragraph{Refinement of the typical distance:} 
Now, the space consumed by the spurious distances must be allotted, \ref{pre_sumOfSpuriousDistances}, Algorithm \ref{alg_getBestDistance}, lines \ref{alg4_correctionStart}-\ref{alg4_correctionEnd}. The correction term for a single good distance is added to the estimate:
	\begin{align}\label{eq_applyCorrectionTerm}
	d^* &\gets d_{\rm typ} + 
		\min\left(d_{\rm typ}, \frac{\sum\limits_{j}d_{{\rm s},j}}{numO\!f\!GoodDistances }\right)
	\;.
	\end{align}
The sum of all spurious distances has to be distributed to all good distances.
Since the derived typical distance cannot be smaller than a half of the true distance, the correction term is limited to the value of $d_{\rm typ}$.
\paragraph{Final estimate:} 
Finally, we can set the parameter estimate to
	\begin{align}\label{eq_finale_a3}
		\hat{a}_3 \gets  \pi / d^* \qquad (~ = 2\pi\cdot f = 2\pi/T = \pi / (T\cdot 0.5) ~)
		\;.
	\end{align}
This entire heuristic is quite robust and is able to identify a suitable distance
representing the frequency of even very noisy cosine functions.
Notwithstanding, it can fail in extreme cases and ill-posed settings.
\paragraph{Case of a single crossing:} 
The above discussed method can be used if at least two crossings have been found.
When only a single crossing has been detected, the observations represent merely a fraction of one oscillation, \Figu{fig_cosFrac}.
In this case, the available range $x_1 \dots x_N$ must serve as an estimate of the typical distance.
The maximum range is defined by one period of a sine wave, where the curve crosses the mean only in the middle:
	\begin{align*}
		0 < x_N - x_1 < 2\cdot d_0
		\;.
	\end{align*}
The best guess we can made is assuming $x_N - x_1 \approx d_0$, i.e. $\hat{a}_3 = \pi/(x_N-x_1)$, Algorithm \ref{alg_getBestDistance}, lines 34-35.
\paragraph{Final remarks:} 
There are some presumptions that do not have been discussed yet.
The number of spurious distances (even number, \ref{pre_spuriousPairs}) has been considered to influence the calculation of the correction term. However, if the number is odd, it seems not  to be decidable, whether one distance should be excluded or one should be included.

Incorporating presumption \ref{pre_extremalDistance} turned out to be unreliable in cases where less than two periods of the cosine function are contained. Especially in presence of strong noise, the distance between the extremal point positions can vary a lot.

Statement \ref{pre_goodVSspurious} has been indirectly used by the sorting the distance classes spurious $<$ good.
\subsection{Estimation of the phase shift} 
Estimating the phase shift $a_4$ is much less critical than estimating the frequency, but still important. Its domain of definition is limited to $0 \dots 2\pi$.
In principle, one could argue that the cosine function reaches its maximum value when the argument of the cosine is zero, $a_3\cdot x + a_4 =0$, and has its minimum at $a_3\cdot x + a_4 =\pi$.
\Algo{alg_findPhase} shows the required computations.
\begin{algorithm}
	\captionsetup{style=mystyle}
  \caption{\label{alg_findPhase}Estimation of the phase shift}
	  \hfil\begin{minipage}[t]{0.79\textwidth}
	\footnotesize
		\begin{algorithmic}[1]
		\LComment {input: $x_i \dots$ chosen conditions $\quad \lfloor N/3\rfloor \le i\le \lceil 2N/3\rceil$}
		\LComment {input: $y_i \dots$ recorded observations}
		\LComment {input: $\hat{a}_1 \dots $ estimated mean of cosine function}
		\LComment {output: $\hat{a}_4 \dots $ estimated phase shift of cosine function}
		\State {$k \gets \argmax\limits_i(y_i)$ \quad $l \gets \argmin\limits_i(y_i)$}
		\State {$yMax \gets y_k$ \quad $yMin \gets y_l$}
		\If {$(yMax - \hat{a}_1) > (\hat{a}_1 - yMin)$} 
				\State {$\hat{a}_4 \gets (-\hat{a}_3 \cdot x_k)$}
		\Else
				\State {$\hat{a}_4 \gets \pi -\hat{a}_3 \cdot x_l$}
		\EndIf
	\end{algorithmic}
	\end{minipage}
\end{algorithm}
The idea behind this procedure is to align the peak of one half-wave with a strong extremum. This extreme should ideally be located in the middle of the signal. If the phase alignment is placed at the border of the sampled signal, a too less accurate frequency estimate can lead to an unfavourable phase shift at the other end of the signal. And curve fitting using this initial parameter vector could miss the valley containing the global minimum. So, if sufficient signal values are available ($\ge 10$) and at least about three periods of the wave-form are contained, the search for extrema should be restricted to the inner third of the signal. These threshold guarantee that the inner third comprises at least one well-sampled period.
\section{Example}
\begin{figure}
	\centering
\ifTIKZinline
	\input{./TIKZ-Plots/cosExample_noise30_P10_fs20.tex}
\else
	\includegraphics[width=\textwidth]{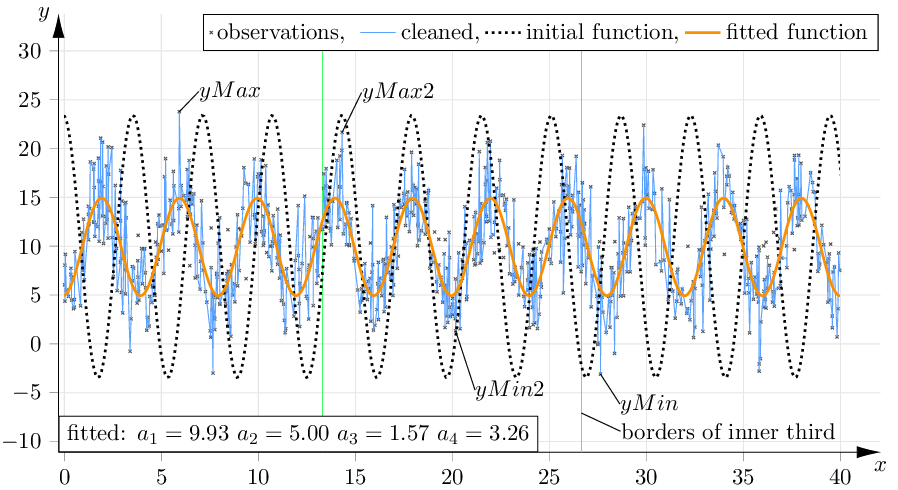}
\fi
	\caption{\label{fig_cosExample}Example of a sampled trigonometric function with certain parameters $a_1 \dots a_4$, the estimated function using the proposed approach, and the fitting result}
\end{figure}

Let us assume that 800 data points $(x_i;y_i)$ are given according to \Figu{fig_cosExample}.
The signal comprises ten periods of a cosine with a frequency of $f=0.250$.
Applying the proposed method, the mean of all observations is $\hat{a}_1 \approx 9.9714$ and the extreme points are
\begin{align*}
	 yMin: (x,y) \approx (27.625,-3.090) ;\quad\mbox{and}\quad 
	 yMax: (x,y) \approx (5.924,23.768)
\end{align*}
leading to $\hat{a}_2 \approx 0.5\cdot[23.768 - (-3.090)] \approx 13.429$.
The spike removal, Algorithm \ref{alg_removeSpikes}, detects and filters out 43 spikes. These points are not connected with the blue curve in Figure \ref{fig_cosExample}.

\begin{figure}
	\centering
\ifTIKZinline
	\input{Example/cos_noise30_P10_fs0020distances.tex}
\else
	\includegraphics[width=15.5cm]{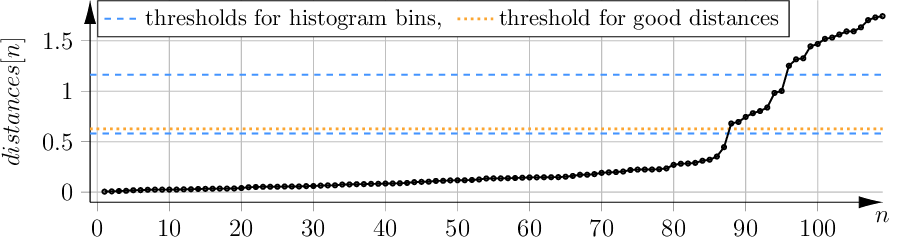}
\fi
	\caption{\label{fig_cosExampleDistances}Detected distances for example function of Figure \ref{fig_cosExample}}
\end{figure}

Following Algorithm \ref{alg_findCrossings}, 110 crossings have been detected. \Figu{fig_cosExampleDistances} shows the corresponding $numDists=109$ distances in sorted order.
Since the largest distance is larger than two times the shortest, the region separating the two classes of distances must be determined. At first, the reference distances has to be defined, Algorithms \ref{alg_findReferenceDistance} and \ref{alg_findReferenceDistancePart2}.
When the distance range is separated into three parts with the thresholds $0.581822, 1.163064$, the middle histogram bin contains eight and the upper bin 14 distances. Both bins fulfill the presumption \ref{pre_noSpurious} but merging both bins does not satisfy this condition. So, the median distance from both bins is taken as reference distance:	$d_{\rm ref} \approx 1.4445$.

There is nothing suspicious about this value and the threshold $d_{\rm ref} / 2.3$ can directly be applied leading to  smallest good distance equal to $distances[88]=0.680176$. The number of good distances is $109-88+1=22$, which is unfortunately larger than the actual number of 20. Therefore, two short and spurious distances influence the computation of the median $d_{\rm typ,median}$ and the mean value $d_{\rm typ,mean}$.
The typical distance is eventually equal to
\begin{align*}
	d_{\rm typ} \approx 0.5\cdot (d_{\rm typ,median} + d_{\rm typ,mean})\approx 0.5\cdot (1.38529 + 1.27032) \approx 1.3278
	\;.
\end{align*}

The sum of the remaining 87 distances is computed and the correction term derived:
\begin{align*}
	correctionTerm &= 	\min\left(d_{\rm typ}, 
			\frac{\sum\limits_{j=1}^{87} d_{{\rm s},j}}{numO\!f\!GoodDistances}\right)  \\
			&\approx \min\left(1.3278, \frac{10.310223}{22}\right)
			\approx 0.4686
	\;.
\end{align*}
Also this correction term is too low, because the sum misses two distances and the denominator is to high.
The final estimated distance is therefore: $d^* \approx 1.3278 + 0.4686 \approx 1.7964$. 
According to (\ref{eq_finale_a3}) the frequency parameter is estimated as
	$%
		\hat{a}_3 \gets  \pi / d^* \approx 1.749
	$, %
which corresponds to a frequency of $\hat{f}= \hat{a}_3 /(2\pi) \approx 0.278$. This value deviates by more than 10\% from the true frequency of $0.250$.

The phase-shift estimation looks for the strongest deviation from the estimated mean.
One finds that 
$$
|yMax2-\hat{a}_1| \approx |21.664 - 9.9714| \approx 11.950 \quad > \quad 
|yMin2-\hat{a}_1| \approx |1.3278 - 9.9714| \approx 8.643
$$
and we take the maximum observation as reference, Algorithm \ref{alg_findPhase}:
$$
 \hat{a}_4 \gets - \hat{a}_3 \cdot x_k
	\approx -25.031 (+8\pi) \approx 0.1013
	\;.
$$
In total, the initial parameters are approximately:
$$
\hat{a}_1 \approx 9.971, \quad
\hat{a}_2 \approx 13.43, \quad
\hat{a}_3 \approx 1.749, \quad
\hat{a}_4 \approx 0.1013
\;.
$$
The corresponding initial function graph is plotted in Figure \ref{fig_cosExample}. It an be seen that the phase of the cosine is aligned with the $x$-position of $yMax2$.
Although the frequency could not be estimated with high accuracy, a correct nonlinear curve fit is possible (orange curve).
\section{Validation of initial parameter estimation}
\subsection{Implementation}
\subsubsection{Proposed method}
The proposed algorithm for initial parameter estimation has been implemented in ANSI-C as extension of the software that was provided together with \cite{Stru16}.
The nonlinear optimization is performed using the Levenberg-Marquardt method according to
\begin{align*}
	\Delta\hat{\mathbf{a}} &= 
	\left(\mathbf{J}^{\rm T}\cdot \mathbf{J} + \mu\cdot \mathbf{I}\right) ^{-1}
		\cdot \mathbf{J}^{\rm T}\cdot \mathbf{r}
		\qquad 
		\begin{array}{l}
		\mathbf{J} \dots \mbox{Jacobian matrix}, \\
		\mathbf{I} \dots \mbox{identity matrix}, \\
		\mathbf{r}=\left(y_i - f(\ul{x}_i|\hat{\ul{a}})\vphantom{W_W^W}\right) \dots  \mbox{vector of residues}
		\end{array}
	\;,
\end{align*}
where the matrix inversion is computed using the singular value decomposition (SVD).
The value of $\mu$ is set initially to 0.001 times the maximum diagonal element of the normal matrix. After successful optimization steps it is reduced by factor 0.125 and after unsuccessful steps it is increased with factor 9.0.
The iterations stops when the (absolute) update steps for all parameters have fallen below $10^{-13}$.
\subsubsection{Lomb-Scargle periodogram}
For comparison purposes, the frequency is additionally estimated based on a Lomb-Scargle periodogram. 
Alternatively to the proposed frequency estimation, equations (\ref{eq_LombScargleP}) and  (\ref{eq_LombScarglePhi}) have been implemented. The range and number of different frequencies to be tested are defined with
\begin{align}\label{eq_fGridLS}
	f_{\min} &= 0.5\cdot \frac{1}{T}  \qquad \mbox{half of smallest detectable frequency} \nonumber\\
	f_{\max} &= \frac{N}{T}  \qquad \mbox{maximum frequency for equidistant samples} \\
	\Delta f &= \frac{1}{5\cdot T}   \qquad \mbox{fivefold oversampling}  \nonumber
\end{align}
and $T = x_N - x_1$.
Therefore, the spectral power $p(\omega)=p(2\pi\cdot f)$ is calculated $5N$ times.
The upper frequency limit $f_{\max}$ is set relatively low. Some sources suggest to use the reciprocal of the minimum distance between the unevenly sampled points. However, when allowing such high frequencies, the optimization procedure tend to overfitting for short signal snippets with only few data points.
The minimum frequency is set to the half of the theoretically smallest detectable frequency in order to include the target frequency also for signal snippets with $P\le 1$.

The frequency with the highest response is taken as the most likely value:
\begin{align*}
	f_{\rm peak} = \argmax_f[p(2\pi\cdot f)]
\end{align*}
All other computations are identical to the proposed method.
\subsection{Investigations and Results}
The effectiveness of the proposed estimation scheme is investigated dependent on three parameters: 
\begin{itemize}
	\item the strength of noise in terms of the signal-to-noise-ratio (SNR)
		\begin{align}
			SNR[dB] = 10 \cdot \log_{10}\left(\frac{\sigma_x^2}{\sigma_z^2}\right) 
			\;,
		\end{align}
	\item the number of covered periods $P$ of the trigonometric wave, and 
	\item the sampling frequency $f_{\rm s}$.
\end{itemize}
The SNR is calculated based on the theoretical variance of a cosine function with amplitude $a_2$, which is $\sigma^2_x=0.5\cdot a_2^2$, and based on the variance $\sigma_z^2$ of the generated noise.
	
As object of investigation, the function 
\begin{align}\label{eq_cosFuncTest}
	y = 10 + 5\cdot \cos(2\pi\cdot f\cdot x + 1) \qquad\mbox{with}\quad f = 0.25 \;(\leadsto a_3 \approx 1.5708)
	\;,
\end{align}
is chosen. This function has been unevenly sampled, while the resulting number of signal values depends on the sampling duration and the sampling frequency. After sampling, normally distributed noise with varying variance has been added.

\subsubsection{Results of the proposed method}
The effectiveness of the proposed approach is simply measured on whether the nonlinear optimization technique can find the global minimum in the error landscape or not. This depends mainly on how accurate the frequency of the model function can be initially estimated.

The Tables \ref{tab_resultsFixP10} - \ref{tab_resultsFixP2} contain the initial frequencies before and final frequencies after application of the nonlinear method of least-squares. Each Table corresponds to a certain signal length in terms of covered periods of a cosine function. \Table{tab_resultsFixP10} shows the results when the measurement includes $P=10$ periods.
	\begin{table}
	\caption{\label{tab_resultsFixP10}Results of curve fitting with fixed number of $P=10$ periods covered\vphantom{$W_W$}. See text for details.}
	\begin{center}\tabcolsep2pt
	\begin{tabular}{|c|cccccc|}
			\hline
			\diagbox{SNR}{$f_{\rm s}$}	
						& 2				&		2.5  			 &     5   			 &    10  		 &     20  		 &     40 		\\
			\hline                                           
23.0 	&	0.250/0.250  & 0.252/0.250  & 0.250/0.250  & 0.250/0.250  & 0.250/0.250  & 0.250/0.250		\\%
17.0 	&	0.250/0.250  & 0.250/0.250  & 0.250/0.250  & 0.251/0.250  & 0.251/0.250  & 0.250/0.250		\\%
11.0 	&	0.250/0.249  & 0.245/0.250  & 0.251/0.250  & 0.250/0.250  & 0.250/0.250  & 0.250/0.250		\\%
7.4 	&	0.250/0.251  & 0.241/0.251  & 0.250/0.250  & 0.252/0.250  & 0.251/0.250  & 0.249/0.250		\\%
4.9 	&	0.245/0.250  & 0.258/0.251  & 0.260/0.250  & 0.252/0.251  & 0.255/0.250  & 0.247/0.250		\\%
3.0 	&	0.231/0.252 &{\it 0.224}/0.250& 0.252/0.249& 0.248/0.249  & 0.253/0.250  & 0.257/0.250 		\\%
1.4 	&	0.239/0.250  & 0.243/0.249  & 0.259/0.251  & 0.234/0.249&{\it 0.278}/0.250& 0.251/0.250	  \\%
0.09	&	0.256/0.252  & 0.239/0.247  & 0.261/0.251  & 0.253/0.249&{\it 0.278}/0.251& 0.272/0.250  \\%
			\hline
	\end{tabular}
	\end{center}
	\end{table}
Frequencies deviating more than about 10\% from the true value are printed in italics. 
Each column corresponds to a certain sampling frequency. Since the test signal has a frequency of $f=0.25$, a sampling frequency of $f_{\rm s}=2.5$ means that ten samples per period have been taken on average.
The signal-to-noise ratio is row-wise gradually reduced. 
At all test points, Levenberg-Marquardt optimization can successfully fit a curve to the noisy observation.

When the sampling time is halved and only five periods are contained, the behaviour of the proposed approach is about the same, \Table{tab_resultsFixP5}.
	\begin{table}
	\caption{\label{tab_resultsFixP5}Results of curve fitting with fixed number of $P=5$ periods covered\vphantom{$W_W$}.  See text for details.}
	\begin{center}\tabcolsep2pt
	\begin{tabular}{|c|cccccc|}		
			\hline
			\diagbox{SNR}{$f_{\rm s}$}	
						&  2 			&	2.5   			&     5  		  &    10 			  &     20   &     40 		\\
			\hline                                           
23.0 	& 0.248/0.250  & 0.253/0.250  & 0.246/0.250  & 0.249/0.250  & 0.248/0.250  & 0.250/0.250 \\%
17.0 	& 0.250/0.250  & 0.252/0.250  & 0.246/0.249  & 0.249/0.250  & 0.247/0.250  & 0.249/0.250 \\%
11.0 	& 0.253/0.250  & 0.253/0.249  & 0.251/0.250  & 0.246/0.249  & 0.249/0.250  & 0.250/0.250 \\%
7.4 	& 0.260/0.250  & 0.230/0.252  & 0.252/0.247  & 0.248/0.250  & 0.248/0.250  & 0.243/0.250 \\%
4.9 	& 0.263/0.256 &{\it 0.218}/0.243& 0.244/0.250 & 0.243/0.249 	& 0.264/0.249  & 0.239/0.250 \\%
3.0 	& 0.255/0.247  & 0.242/0.247  & 0.232/0.248  & 0.249/0.249  & 0.240/0.249  & 0.238/0.249 \\%
1.4 	& 0.248/0.251  & 0.230/0.249  & 0.242/0.250  & 0.256/0.254  & 0.263/0.251  &{\it 0.305}/0.251   \\%
0.09 	&{\bf	0.210/0.090}&0.266/0.250& 0.243/0.251&{\it 0.278}/0.248&0.268/0.252	&{\bf 0.323/0.320}    \\%
			\hline
	\end{tabular}
	\end{center}
	\end{table}
The initial frequencies now deviate somewhat more strongly from the actual value $f=0.25$. However, there are only two failure cases (printed in bold). Additionally, frequencies differing more than about 10\% are printed in italics. 
At the sampling frequency $f_{\rm s}=2.5$ and $SNR=4.9$dB, the initial frequency estimate is very low. However, the estimated parameter vector is still accurate enough to enable a good fit. Even the very high frequency at $f_{\rm s}=40$ and $SNR=1.4$dB still allows to find the global optimum. Here, the lower number of covered periods is an advantage, because the increasing phase shift between the estimated curve and the observations towards the ends does not reach a critical amount yet.

\begin{figure}
	\centering
\ifTIKZinline
	\input{./TIKZ-Plots/cosExample_noise35_P05_fs02.tex}
\else
	\includegraphics{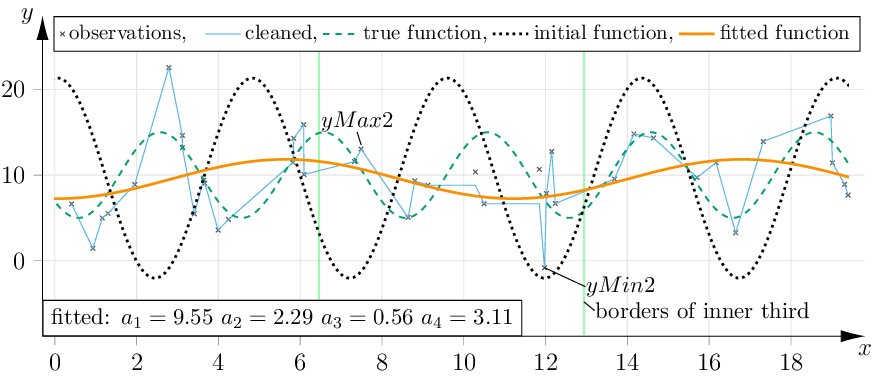}
\fi
	\caption{\label{fig_cosExample_noise35_P05_fs02}Example signal of fitting failure at $P=5$, $SNR=0.09$dB, $f_{\rm s} = 2.$}
\end{figure}

The proposed scheme fails to identify the correct parameters for $f_{\rm s}=2$ and $SNR=0.09$dB. Even by eye, it is almost impossible to make a guess, since the sampling frequency is too low given the high level of noise, \Figu{fig_cosExample_noise35_P05_fs02}.

Further reduction of the measured signal length makes it more difficult to the proposed scheme to determine a signal frequency close to the correct one, \Table{tab_resultsFixP2}.
	\begin{table}
	\caption{\label{tab_resultsFixP2}Results of curve fitting with fixed number of $P=2$ periods covered\vphantom{$W_W$}. See text for details.}
	\begin{center}\tabcolsep2pt
	\begin{tabular}{|c|cccccc|}		
			\hline
			\diagbox{SNR}{$f_{\rm s}$}	
						& 	2					&	2.5   					&     5   			 &    10   					&     20   		&     40 		\\
			\hline                                           
23.0 	& 0.251/0.251  			& 0.274/0.250  		& 0.235/0.250			& 0.255/0.249  		& 0.251/0.250  & 0.251/0.250		\\%
17.0 	& 0.250/0.248  			& 0.265/0.245  		& 0.242/0.246  		& 0.250/0.251  		& 0.253/0.249  & 0.253/0.250		\\%
11.0 	& 0.250/0.256				& 0.263/0.250			& 0.241/0.250  		& 0.248/0.247  		& 0.246/0.249  & 0.253/0.248		\\%
7.4 	& 0.254/0.248  			&{\it 0.349}/0.249& 0.255/0.249  		& 0.249/0.243  		& 0.242/0.248  & 0.259/0.249		 \\%
4.9 	&{\it 0.275}/0.251 	&{\it 0.280}/0.247& 0.233/0.260			& 0.249/0.243  		& 0.240/0.247  & 0.244/0.250		 \\%
3.0 	&{\it 0.176}/0.270	& 0.260/0.251  		& 0.228/0.259			& 0.256/0.253			& 0.255/0.250  & 0.238/0.244  \\%
1.4 	&{\it 0.228}/0.250	&{\it 0.212}/0.231& 0.232/0.240  		& 0.252/0.255  	&{\bf 0.774/0.551}&{\it 0.202}/0.250   \\%
0.09 	&{\bf 0.279}/0.272 	&{\bf 0.171}/0.242&{\bf 0.140/0.001}&{\it 0.347}/0.253&{\bf 0.381/0.534}&{\it 0.402}/0.250    \\%
			\hline
	\end{tabular}
	\end{center}
	\end{table}
Nevertheless, down to an SNR of 3.0 dB, the estimates are sufficiently accurate to fit a reasonable model to the observations.
Stronger noise reduces the chance that enough good distances remain; smaller segments mislead the algorithm.
\subsubsection{Results compared to the Lomb-Scargle periodogram}
Frequency determination using the Lomb-Scargle periodogram is a reliable method, as all data points are evaluated together. The most interesting aspect of comparing the proposed method with this standard technique is to find out whether it is generally possible to find an appropriate frequency value for a given data set.
Therefore, only the failure cases from Tables \ref{tab_resultsFixP10} -- \ref{tab_resultsFixP2} and cases with final frequency deviating more than usual are investigated, \Table{tab_resultsLombScargle}. 
	\begin{table}
	\caption{\label{tab_resultsLombScargle}Results of the Lomb-Scargle method in terms of detected frequency in comparison to the proposed method. Estimated frequencies $\hat{f}$ not being suitable for reasonable curve fitting are printed bold-face.}
	\begin{center}\tabcolsep2pt
	\begin{tabular}{|ccc|ccc|}		
			\hline
					& 			&					&\multicolumn{3}{c|}{$\hat{f}$\vphantom{\Huge I}}		\\
			$P$	&	SNR		& $f_{\rm s}$ 	& Lomb-Scargle	&	proposed & fitted 		\\
			\hline                                           
			5		&	0.09	&	2.0			&	0.253		&{\bf	0.198}&	0.255		\\
			5		&	0.09	&	40			&	0.245		&{\bf	0.323}&	0.250		\\
			\hline
			2		&	7.4		&	2.5			&{\bf	0.458}&	0.349		&	0.249		\\
			2		&	3.0		&	2.0			&	0.263		&	0.176			&	0.270		\\
			2		&	1.4		&	2.5			&	0.244		&	0.212			&	0.231		\\
			2		&	1.4		&	20			&	0.252		&{\bf	0.774}&	0.258 \\
			2		&	1.4		&	40			&	0.256		&	0.202			&	0.250 \\
			\hline
			2		&	0.09	&	2.0			&	0.263		&	0.279			&	0.272	\\
			2		&	0.09	&	2.5			&	0.244		&	0.171			&	0.242	\\
			2		&	0.09	&	5.0			&	0.265		&{\bf	0.140}&	0.271		\\
			2		&	0.09	&	10			&	0.264		&{\bf	0.347}&	0.253	\\
			2		&	0.09	&	20			&	0.252		&{\bf	0.381}&	0.247	\\
			\hline
	\end{tabular}
	\end{center}
	\end{table}
Interestingly, also the Lomb-Scargle periodogram failed to identify the correct frequency at one working point: $P=2$, $SNR=7.4$dB, $f_{\rm s}=2.5$. 
The values printed in bold do not lead to a correct fitting result.

\begin{figure}
	\centering
\ifTIKZinline
	\begin{minipage}{0.49\textwidth}\centering
		\input{./Multi_P01/cosExample_noise20_P01_fs02.tex}\\ %
		(a)
	\end{minipage}
	\hfil 
	\begin{minipage}{0.49\textwidth}\centering
		\input{./Multi_P01/cosExample_noise25_P01_fs02.tex}\\ %
		(b)
	\end{minipage}

	\hfil 
	\begin{minipage}{0.49\textwidth}\centering
		\input{./Multi_P01/cosExample_noise25_P01_fs03.tex} \\ %
		(c)
	\end{minipage}
	\hfil 
	\begin{minipage}{0.49\textwidth}\centering
		\input{./Multi_P01/cosExample_noise25_P01_fs05.tex} \\ %
		(d)
	\end{minipage}

	\hfil 
	\begin{minipage}{0.49\textwidth}\centering
		\input{./Multi_P01/cosExample_noise30_P01_fs02.tex}\\ %
		(e)
	\end{minipage}
	\hfil 
	\begin{minipage}{0.49\textwidth}\centering
		\input{./Multi_P01/cosExample_noise30_P01_fs03.tex}\\ %
		(f)
	\end{minipage}

	\hfil 
	\begin{minipage}{0.49\textwidth}\centering
		\input{./Multi_P01/cosExample_noise35_P01_fs03.tex} \\ %
		(g)
	\end{minipage}
	\hfil 
	\begin{minipage}{0.49\textwidth}\centering
		\input{./Multi_P01/cosExample_noise35_P01_fs20.tex} \\
		(h)
	\end{minipage}
\else
	\includegraphics{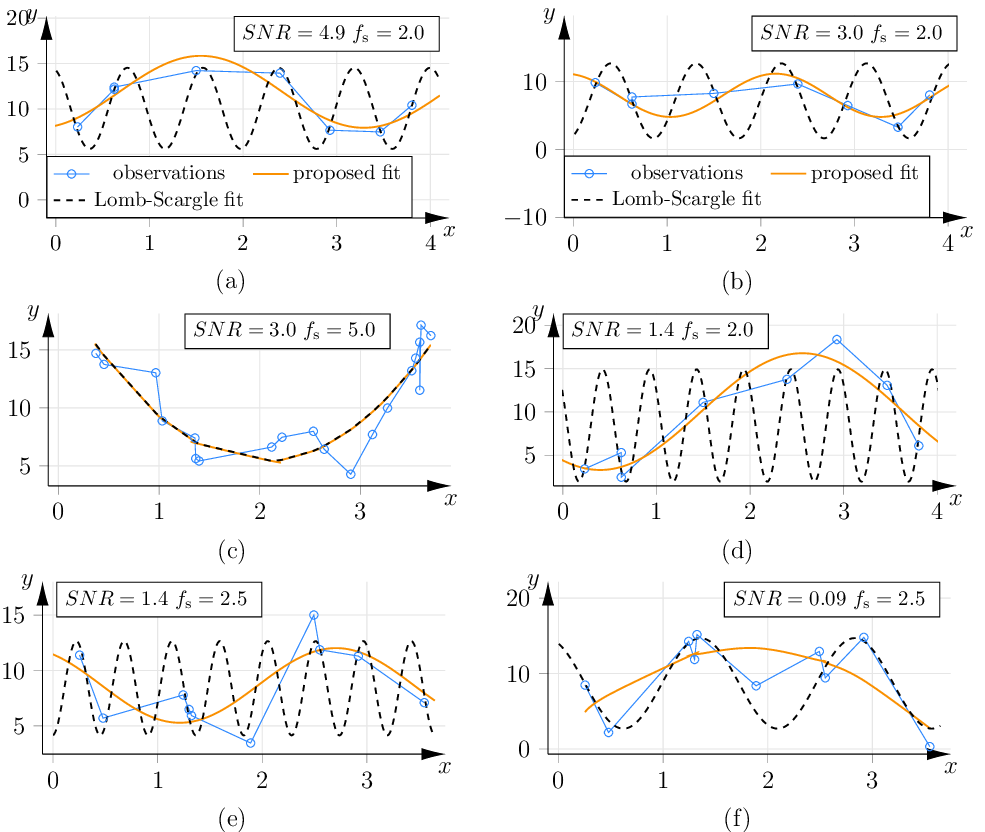}
\fi
	\caption{\label{fig_cosExample_comparison}Comparison of fitting results with $P=1$}
\end{figure}

\Table{tab_resultsFixP1} proves that the proposed method can deal with very short signal samples. %
	\begin{table}
	\caption{\label{tab_resultsFixP1}Comparison of Lomb-Scargle and proposed method for signal snippets covering only $P=1$ period\vphantom{$W_W$}. The values pairs show frequency values after least-squares fits based on Lomb-Scargle estimates / FIPEFT estimates.}
	\begin{center}\tabcolsep2pt
	\begin{tabular}{|c|cccccc|}		
			\hline
			\diagbox{SNR}{$f_{\rm s}$}	
				&  2						& 2.5   		&     5    			&    10  			 		&     20  			 &     40 		\\
			\hline                                           
23.0 	& 0.253/0.253   & 0.252/0.252   	& 0.254/0.254   	& 0.253/0.253   & 0.252/0.252   & 0.254/0.254\\%
17.0 	& 0.245/0.245   & 0.239/0.239   	& 0.265/0.265   	& 0.245/0.245   & 0.255/0.255   & 0.250/0.250\\%
11.0 	& 0.237/0.237   & 0.236/0.236   	& 0.300/0.300   	& 0.242/0.242   & 0.229/0.229   & 0.244/0.244\\%
7.4 	& 0.246/0.246   & 0.261/0.261   	& 0.218/0.218   	& 0.260/0.260   & 0.247/0.247   & 0.252/0.252\\%
4.9 	&{\bf	2.223}/0.287& 0.329/0.329   & 0.272/0.272   	& 0.217/0.217   & 0.265/0.265   & 0.252/0.252\\%
3.0 	&{\bf 1.098}/0.445&	0.201/0.201		&{\it 0.002/0.002}& 0.252/0.252   & 0.260/0.260   & 0.259/0.259 \\%
1.4 	&{\bf	1.988}/0.232&{\bf 2.187}/0.335& 0.265/0.265		& 0.281/0.281 	& 0.253/0.253   & 0.275/0.275 \\%
0.09 	&	0.277/0.277		&{\bf 0.679/0.001}& 0.276/0.276   	& 0.192/0.192   & 0.235/0.235		& 0.268/0.268     \\%
			\hline
	\end{tabular}
	\end{center}
	\end{table}
In contrast to previous tables, the left value in each cell now shows the fitting result based on the Lomb-Scargle estimates. The right value shows, as before, the fitting result based on the initial frequency determined by FIPEFT.
In five cases (printed in bold), the methods produce strongly deviating results, mostly because there are too few observations for a distinct picture. 
In one case, printed in italics, both methods end up with a suspicious low frequency.
\Figu{fig_cosExample_comparison} shows the corresponding plots.
The FIPEFT method generally tends to choose a low-frequency fit, while the Lomb-Scargle approach looks for a global best fit. 
Since based on (\ref{eq_fGridLS}) the maximum frequency tested for the Lomb-Scargle periodogram can be higher than half of the actual sampling frequency, this method can pick an aliasing frequency, plots (a), (b), (d), and (e).

In sub-plot (c), the recorded data appear as one half-wave instead of a full period of oscillation. Both methods produce a nice looking fit. However, the resulting frequency is only $0.002$. The shape of the fitted curve results from extremely high values for $\hat{a}_1$ and $\hat{a}_2$. This also holds true for plot (f). Although the curve based on the FIPEFT makes a good visual impression, the Levenberg-Marquardt optimization finds the best fit at about $\mathbf{a}=(-95927,\; 95941,\; 0.0086,\; 6.268)$.

\begin{figure}
	\centering 
\ifTIKZinline
	\begin{minipage}{0.49\textwidth}\centering
		\input{./Multi_P00/cosExample_noise10_P00_fs03.tex}\\
		(a)
	\end{minipage}
	\hfil 
	\begin{minipage}{0.49\textwidth}\centering
		\input{./Multi_P00/cosExample_noise20_P00_fs05.tex}\\
		(b)
	\end{minipage}

	\hfil 
	\begin{minipage}{0.49\textwidth}\centering
			\input{./Multi_P00/cosExample_noise20_P00_fs40.tex}\\
		(c)
	\end{minipage}
	\hfil 
	\begin{minipage}{0.49\textwidth}\centering
		\input{./Multi_P00/cosExample_noise25_P00_fs05.tex} \\
		(d)
	\end{minipage}

	\hfil 
	\begin{minipage}{0.49\textwidth}\centering
		\input{./Multi_P00/cosExample_noise35_P00_fs10.tex}\\
		(e)
	\end{minipage}
	\hfil 
	\begin{minipage}{0.49\textwidth}\centering
		\input{./Multi_P00/cosExample_noise35_P00_fs20.tex}\\
		(f)
	\end{minipage}
\else
	\includegraphics{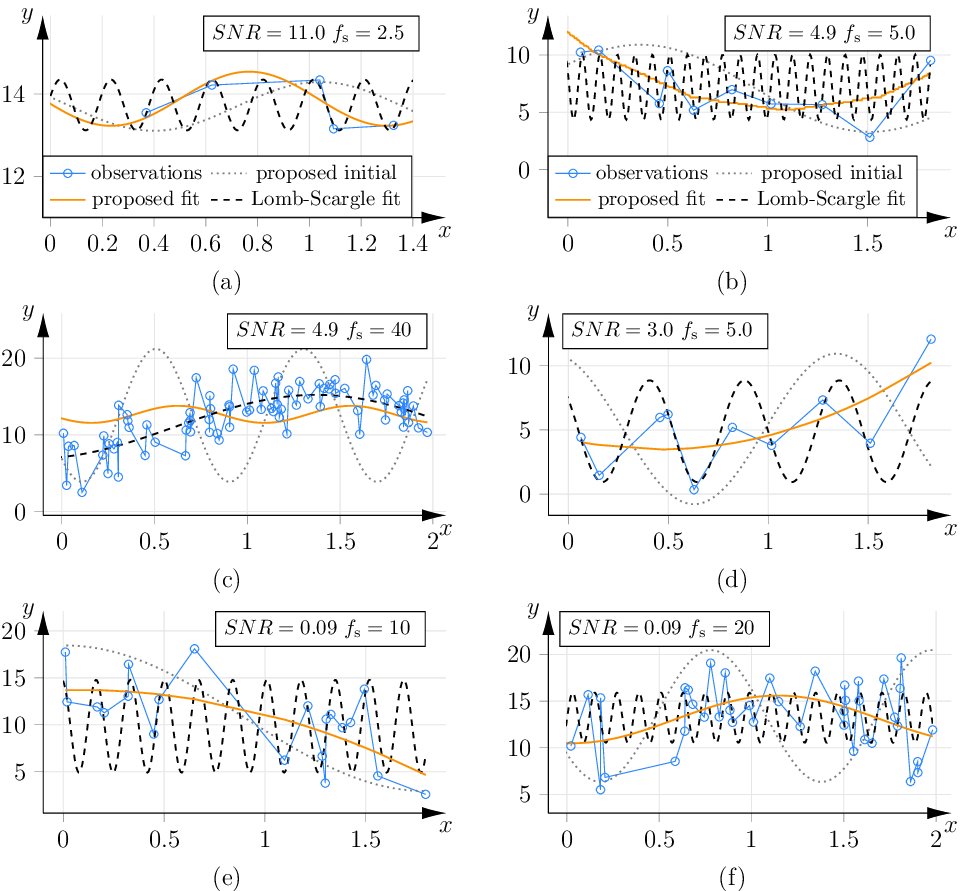}
\fi
	\caption{\label{fig_cosExample_comparisonP00}Comparison of fitting results with $P=0.5$}
\end{figure}

Remembering the goal of the research, also the results for signals which cover less than one period must be investigated. Obviously, it is impossible to make any correct estimate of the frequency in such cases. However, the primary goal is to find an initial parameter vector that enables nonlinear fitting procedures to match the observations to a reasonable curve. Still we use the frequency as indicator for putatively failures, \Table{tab_resultsFixP00}.
	\begin{table}
	\caption{\label{tab_resultsFixP00}Comparison of Lomb-Scargle and proposed method for signal snippets covering only $P=0.5$ periods\vphantom{$W_W$}. The values pairs show frequency values after least-squares fits based on Lomb-Scargle estimates / FIPEFT estimates.}
	\begin{center}\tabcolsep2pt
	\begin{tabular}{|c|ccccc|}		
			\hline
			\diagbox{SNR}{$f_{\rm s}$}	
						& 2.5   			&     5    		&    10   			&     20   		&     40 		\\
			\hline                                           
23.0 	& 0.271/0.271  			& 0.281/0.281  & 0.241/0.241  	& 0.222/0.222		& 0.260/0.260		\\%
17.0 	& 0.003/0.003 			& 0.272/0.272  & 0.264/0.264  	& 0.298/0.298 	& 0.180/0.180		\\%
11.0 	&{\bf 5.107}/0.939  & 0.175/0.175  & 0.246/0.246  	& 0.202/0.202 	& 0.256/0.256    \\%
7.4 	&{\bf 3.790}/0.450  & 0.002/0.002  & 0.081/0.081  	& 0.192/0.192 	& 0.307/0.307  \\%
4.9 	&{\bf 4.517}/0.005	&{\bf 4.645}/0.006& 0.455/0.455 & 0.295/0.295 	&	0.330/{\bf 1.255}    \\%
3.0 	&{\bf 4.887}/0.378  &{\bf 2.108/0.002}& 0.005/0.005	& 0.002/0.002 	& 0.350/0.350    \\%
1.4 	&{\bf 1.190/1.190}	&{\bf 3.358/1.160}& 0.352/0.352  & 0.575/0.575	& 0.004/0.004     \\%
0.09	&{\bf	5.071}/0.820  & 0.002/0.992  &{\bf 5.898}/0.002&{\bf 8.355}/0.449& 0.003/{\bf 1.512}			\\%
			\hline
	\end{tabular}
	\end{center}
	\end{table}
The values printed in bold mark the misses. 
\Figu{fig_cosExample_comparisonP00} visualizes six results from Table \ref{tab_resultsFixP00}.
FIPEFT produces reasonable curves for cases (b), (d), (e), and (f), the Lomb-Scargle method only for (c).
In all other cases, both approaches overestimate the frequency and lead to questionable results. 
In general, the best fit in terms of least squares can be found using the periodogram method. However, the high frequency values indicate that the results are related to overfitting and these fits have no predictive capability. 
\subsubsection{Results for real data sets}
\begin{figure}
	\centering
	\includegraphics{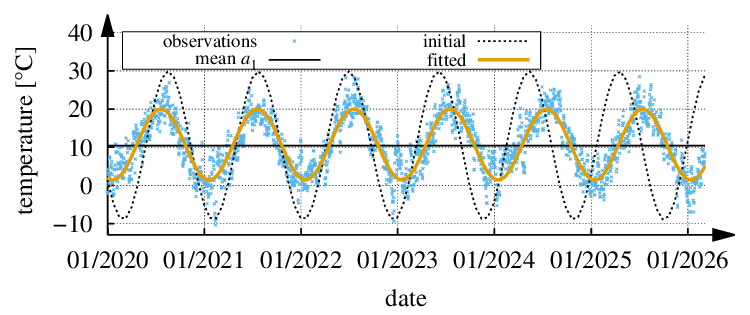} \\
		(a)
	
	\includegraphics{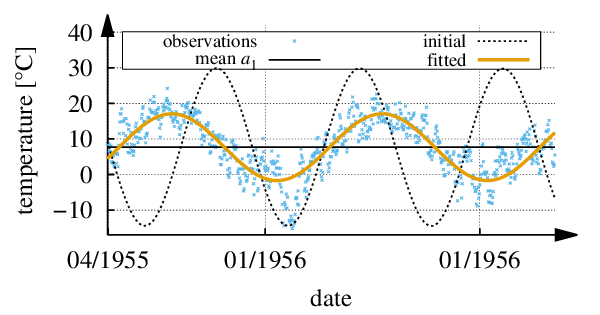}\\
	(b)
	\caption{\label{fig_temperatureNuremberg}Daily average temperature in Nuremberg - results of fitting temperature data: (a) 01.01.2020 -- 06.03.2026, (b) 07.04.1955 -- 09.05.1957}
\end{figure}

The proposed method also has been tested in application to real-world data taken from \cite{DWD26}.
The data comprises the average temperature in Nuremberg, Germany, over a time span of more than six years and about two years, respectively, see \Figu{fig_temperatureNuremberg}.
The data points have arbitrarily been chosen.
The estimated initial frequency for Figure \ref{fig_temperatureNuremberg}(a) corresponds to a period duration of $T=1/f \approx 340.7$. The duration after curve fitting is $T \approx 364.0$. The expected duration is slightly higher than 365 (days), since one leap year is involved.
The initial parameter estimation for the shorter data set, Figure \ref{fig_temperatureNuremberg}(b), is less accurate: $T\approx 244.3$. Nevertheless, it is good enough for the fitting procedure which leads to $T\approx 358.8$.
\subsubsection{Time complexity}
The program including the determination of the initial parameter vector was executed on a notebook equipped with an Intel i7-1165G7\@2.80GHz processor. To measure execution times, the \verb+__rdtsc()+ function has been used. It returns the processor's time stamp counter representing the number of clock cycles since the last system reset \cite{Mic24}.
Two time stamps were recorded: one immediately before the parameter estimation begins, and another right after all parameters have been determined. The difference between these two values is taken as the time measurement.

As already pointed out in \cite{Stru25}, the measured execution time does not necessarily correspond to the actual number of CPU instructions executed during the coding process. Two key factors must be considered. First, the CPU may be concurrently handling other background tasks, which can affect timing accuracy. Second, the execution time is influenced by the CPU's clock frequency, which is not constant.
Modern processors can dynamically adjust their clock speed boosting performance during short, intensive workloads. However, under sustained high loads, such as in our simulations, the CPU temperature may rise significantly, triggering thermal throttling. This mechanism reduces the clock frequency to prevent overheating, and it remains in effect until the temperature drops below a safe threshold.
As a result, the CPU frequency can fluctuate frequently, making time-based measurements unreliable indicators of actual computational effort.

To minimize the impact of background processes, all unnecessary applications were closed, and the internet connection was disabled. The issue of fluctuating CPU frequency was addressed by limiting the maximum processor performance to 75\% via the Windows 11 power settings. This adjustment resulted in a nearly constant CPU frequency of approximately 2.0 GHz, as confirmed through the task manager.

As the determination of $\hat{a}_1$, $\hat{a}_2$, and $\hat{a}_4$ is identical for the FIPEFT and Lomb-Scargle approaches, the time difference stems from the frequency estimation, i.e $\hat{a}_3$.
FIPEFT has $\mathcal{O}(N)$ complexity, since the spike removal, the determination of the extrema, and finding the crossings require accessing of all $N$ signal values.
Lomb-Scargle uses the equations (\ref{eq_LombScargleP}) and (\ref{eq_LombScarglePhi}), where the complexity is also defined by the number $N$ of signal values. However, the number of frequencies to be tested is also a function of $N$, Equation (\ref{eq_fGridLS}). Therefore, the big-O complexity of the Lomb-Scargle approach is $\mathcal{O}(N^2)$.

\begin{figure}
	\centering
	\includegraphics{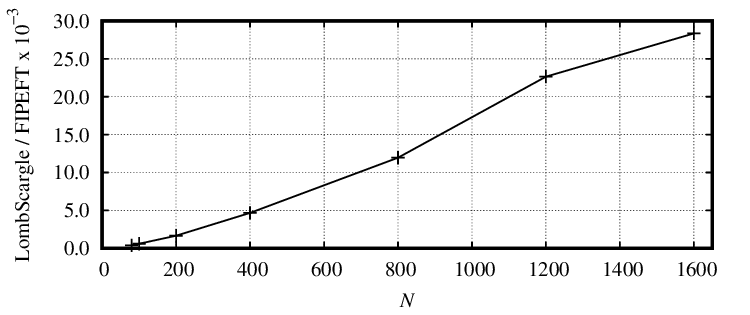}
	\caption{\label{clockCycles_P10}Relation of the measured clock cycles depending on the number of data points}
\end{figure}

The times have been measured for the signals covering ten periods of oscillation sampled with different frequencies and SNRs.
Since the Lomb-Scargle processing time is independent of the strength of noise, the minimum of the measured times at a fixed $f_{\rm s}$ is taken as the reference value. In contrast, the efforts made by FIPEFT may depend on the SNR, since noise affects the number of distances to be processed. For this reason, the median of the time measurements is used as a reference.
\Figu{clockCycles_P10} compares the computational complexity of both approaches using the ratio
	\begin{align}
	\frac{numClockCycles({\rm Lomb{-}Scargle})[N]}{numClockCycles({\rm FIPEFT})[N]}
	\end{align}
as a function of the signal length $N$.
Note that the ratio is $10^3$ times higher than indicated by the vertical scale.
For the shortest considered signal length of $N=80$, the Lomb-Scargle method needs a number of clock cycles that is already about 400 times the number required by FIPEFT. This factor increases with $N$ as can be seen in Figure \ref{clockCycles_P10}. 

\section{Summary and Discussion}
The proposed FIPEFT method is an appropriate low-complex technique to find suitable initial parameter vectors for the nonlinear fitting of unevenly sampled trigonometric functions.
A core element of FIPEFT is the new approach to estimating the frequency of signals.
This component belongs to the family of zero-crossing techniques and reaches its limits when error amplitudes regularly exceed the oscillation amplitude. In this case, half-waves are likely to be divided into shorter parts, and it becomes unlikely that more or less complete half-wave intervals can be found. 
A second important step is the determination of the phase shift that improves the fitting success by aligning the estimated curve with an extremum in the centre of the signal snippet.

In contrast to the Lomb-Scargle periodogram, the proposed method cannot deal with large gaps in recordings because these gaps would be considered as distance candidates. Different from other approaches, the accuracy of FIPEFT results does not improve with increasing signal length, because its capability to alleviate random errors is limited.

FIPEFT is especially robust, when the samples are symmetrically distributed around the mean value of all observations. This maximizes the chance to find good distances not being segmented by large errors in the signal values. In scenarios with low SNR, a relation of sampling frequency $f_{\rm s}$ and signal frequency of about 50:1 seems to maximize the chance of finding a suitable initial frequency. If $f_{\rm s}$ is too low, then the noise can cause missing half-waves, is it too high, the risk of segmented half-waves increases.

Since FIPEFT is extremely fast and requires only a few periods of the waveform, it can also be used for real-time estimations with varying signal parameters. If the data under investigation covers at least about five oscillation periods, the proposed approach is the method of choice down to an SNR of 1.4 dB in systems that require low-complexity processing. 
\section{Acknowledgements}
This text has been type-setted with \LaTeX. The algorithms have been written with the help of package \verb+algpseudocodex+. The figures have been created using \TikZ/PGF  (packages \verb+tikz+ and \verb+pgfplots+) and GnuPlot (version 5.4). The references are included using \BibTeX.
\includegraphics{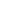}

\bibliography{literature} 
\bibliographystyle{alphaTSnew}
\end{document}